\documentclass[prd,
nofootinbib,
 amsmath,amssymb,
 aps,
]{revtex4-2}
\usepackage{geometry}
\usepackage{amsmath,amssymb}
\usepackage{slashed,verbatim,graphicx}
\usepackage{enumerate}
\usepackage{placeins}
\graphicspath{{/figures/}}
\usepackage{ytableau}
\usepackage{pstricks}
\usepackage{color}
\usepackage{physics}
\usepackage{mathtools}
\usepackage{tikz-cd}
\usetikzlibrary{decorations.pathreplacing,decorations.markings, fit, calc}

\usepackage{hyperref}
\usepackage{float}
\usepackage[caption=false]{subfig}

\newcommand{\nn}{\nonumber \\}
\newcommand{\mZ}{\mathbb{Z}}

\newcommand{\mC}{\mathbb{C}}

\newcommand{\mrm}[1]{\mathrm{#1}}

\newcommand{\mc}{\mathcal}

\DeclareMathOperator{\sh}{sh}

\tikzset{
    on each straight segment/.style={
    decorate,
    decoration={
        show path construction,
        moveto code={},
        lineto code={
            \path [#1]
            (\tikzinputsegmentfirst) -- (\tikzinputsegmentlast);
        },
        curveto code={
            \path  (\tikzinputsegmentfirst)
            .. controls
            (\tikzinputsegmentsupporta) and (\tikzinputsegmentsupportb)
            ..
            (\tikzinputsegmentlast);
        },
        closepath code={
            \path 
            (\tikzinputsegmentfirst) -- (\tikzinputsegmentlast);
        },
    },
},
mid arrow/.style={postaction={decorate,decoration={
            markings,
            mark=at position .6 with {\arrow[#1]{stealth}}
}}},
mid arrowpos/.style 2 args={postaction={decorate,decoration={
            markings,
            mark=at position #2 with {\arrow[#1]{stealth}}
}}},
}

\ytableausetup
{boxsize=0.9em}
\ytableausetup
{aligntableaux=center}

\begin{document}

\title{Spin Chains from large-$N$ QCD at strong coupling}
\author{David Berenstein}
\author{Hiroki Kawai}
\affiliation{Department of Physics, University of California at Santa Barbara, CA 93106}
\date{\today}

\begin{abstract}
    We study the strong coupling expansion of large $N$ QCD in various dimensions, reformulating the Kogut-Susskind Hamiltonian on a square lattice in terms of (constrained) one dimensional spin chain models. We study the integrability properties of the spin chain obtained this way: there is large class of integrable subsectors, but we show that the full spin chain is not integrable, at least when viewed from a description based on Bethe ansatz. We demonstrate that the spin chains no longer possess integrability due to the constraints arising from the zigzag symmetry of the confining strings. 
    The spin chain description properly estimates the roughening transition point by extrapolating the first-order analytical results   based on integrability of some subsectors.   
    The generalization to higher dimensions are also considered, where we also find the small subsectors without the zigzag constraints to be integrable. 
\end{abstract}

\maketitle

\tableofcontents

\section{Introduction}
The recent advances of quantum computing technology open a new avenue to perform numerical simulations of quantum field theories (QFTs). 
These  allow us to simulate, at least in principle, the lattice theory in the Hamiltonian formalism and hence to circumvent the sign problem in the conventional Monte-Carlo methods with Lorentzian actions for simulating real-time dynamics. 

In principle, if one could find the finite-dimensional Hilbert space containing the low-energy modes of the field theory and write the effective Hamiltonian as an operator (i.e. a spin operator for a digital quantum computer with qubits) in this finite Hilbert space, then it should be expected that the observables in interest can be simulated on a digital quantum computer~\cite{Byrnes:2005qx}. 
Still, one has to rigorously check whether the original theory and the effective Hamiltonian are in the same universality class. 
For gauge theories, the Kogut-Susskind formulation of lattice gauge theory Hamiltonian~\cite{Kogut:1974ag} is a good starting point. 
Even though the gauge field Hilbert space is infinite-dimensional in their formalism as an ordinary bosonic field Hilbert space, several approaches have been proposed for deforming it to finite-dimensional. 
Roughly speaking, one has two choices to regularize the Hilbert space: (i) by introducing a cutoff to the electric flux by hand while effectively preserving the Lie algebra structure of the gauge symmetry, based for example on the quantum link model~\cite{Chandrasekharan:1996ih,Brower:1997ha,Beard:1997ic,Brower:1997ha,Brower:2003vy, Brower:2020huh,Berenstein:2022wlm} or the loop-string-hadron framework~\cite{Raychowdhury:2019iki, Kadam:2022ipf}, or (ii) by replacing the gauge symmetry with another that naturally leads to finite-dimensional representations and has the original gauge group as some limit, such as the discrete subgroup~\cite{Zohar:2016iic, Lamm:2019bik} or the q-deformation~\cite{Zache:2023dko, Hayata:2023bgh}. 
In either case, a full simulation of a theory like QCD requires computational resources far beyond of what is available on current devices, especially in dimensions higher than $1+1$-D~\cite{Kan:2021xfc, Bauer:2022hpo, Santra:2025dsm}. 

Soon after the discovery of asymptotic freedom of QCD~\cite{Gross:1973ju,Politzer:1973fx,Gross:1973id,Gross:1974cs}, Wilson beautifully demonstrated that gauge theory in the strong coupling regime can be formulated on lattice spacetime, and the Wilson loop expectation value follows the area law, which indicates quark confinement~\cite{Wilson:1974sk}. 
One can interpret the flux tube in the strong coupling regime as a string-like object with confined quarks at its endpoints, since its energy is exactly proportional to its length as $g_{\text{YM}} \rightarrow \infty$. 
This can be understood in the lattice formulation provided that the flux tube is the gluon excitation on a path connecting the quarks~\cite{Kogut:1974ag}. 
The confining string description becomes even clearer in the large $N$ limit~\cite{tHooft:1973alw}, where the nonplanar interactions of the confining strings are suppressed, corresponding to the genus expansion of string theory. 
Of course, the Lagrangian description of this confining string needs a modification from the pure Nambu-Goto worldsheet action due to its conformal anomaly since it propagates in spacetime with noncritical ($4 \neq 26$) dimensions. 
There have been several proposals for modifications of the confining string action such as~\cite{Polchinski:1991ax,Dubovsky:2015zey}. 

Based on this confining string approach for QCD, we have proposed a potential alternative approach to use quantum computers for lattice gauge theory~\cite{Berenstein:2023lgo} by working in the Hamiltonian formalism of the strong coupling expansion. 
The framework restricts the lattice $SU(N)$ Yang-Mills theory in the large $N$ limit to the sector with a Wilson line or a confining string, where the relevant part of the Kogut-Susskind Hamiltonian action ends up being reduced to a $1+1$-D spin chain Hamiltonian action for the coinfining string.  
Note that the limit taken there is not the 't Hooft limit but a double scaling limit $\lambda, N \rightarrow \infty$ with $\lambda \gg N$ so that the strong coupling expansion makes sense. In this limit, the magnetic terms of the plaquette action are turned off at zeroth order. Instead, one diagonalizes the electric (kinetic) terms first and proceeds from there by utilizing the magnetic terms as a small perturbation. 

In this limit, the Hilbert space near the vacuum is described as a non-interacting gas of strings, and 
the leading order corrections to the one string states  close in the one string sector. The string states can be described as  \textit{words} of \textit{letters} instructing the directions of the link excitations as one moves along the string. This is similar in spirit to the spin chains arising from ${\cal N}=4$ SYM \cite{Berenstein:2002jq}, but here the words actually indicate displacements on the lattice directions by the operator matrices.
The plaquette actions on these states become manipulations of the letters making a word. 
Translating the letters to spin states, it provides a systematic construction of spin chains corresponding to the perturbation corrections from strong-coupling  to the confining string states, generalizing the idea of formulating the perturbation analysis as a free fermion theory in~\cite{Kogut1981}.  
The spin chains discussed in~\cite{Berenstein:2023lgo} were found to be integrable, while those in more general settings--including higher-order corrections, higher spacetime dimensions, or more general sectors--are expected to lose integrability. Part of the intention of this paper is to address instances on where integrability is lost.
In our previous work we showed that integrability would arise to leading order perturbation theory in all dimensions, so long as the zig-zag symmetry constraints were ignored. In that sense, these spin chains are closed to being integrable. Here, we explicitly write the effects of the zig-zag constraints. These require additional projectors in the spin chain to forbid states that are not allowed by the constraints. These projectors change the leading order nearest neighbor interaction to an interaction that includes four neighboring sites instead. We show that these can destroy the integrability of the spin chain, but many subsectors are still integrable.

Notice that because we obtain spin chains where the Hamiltonian is local on the spin chain,  they are suitable for quantum simulation with current devices. 
Remarkably, in the $2+1$-D spacetime, the spin chain Hamiltonian is found to be integrable within a larger set of subsectors than what one would have initially thought. Because of that some exact calculations are accessible at leading order in the perturbation of a large class of interesting states.

The philosophy of this idea is somewhat similar to the recent effort on quantum simulation of quantum gravity using matrix models~\cite{Rinaldi:2021jbg}. 
The matrix models are well-defined quantum mechanical models, so they can be simulated on quantum computers in a straightforward way, even though the full simulation of quantum gravity itself is not feasible with our current understanding of quantum gravity. 

We must accept some tradeoffs for the lower computational cost we are achieving. For a simulation of individual confining strings, it would not allow us to access  QCD in the regime where it would be closest to nature: a strong coupling relativistic theory (the continuum limit of the lattice system). 
In particular, the deconfined phase would be beyond reach. 
The situation is similar for the case of matrix models. For example the BFSS model is conjectured to be dual to M-theory in asymptotically flat backgrounds~\cite{Banks:1996vh}--it is not expected to grasp the whole eleven-dimensional quantum gravity theory especially in the strong-coupling regime where the nonperturbative effects involving saddle-point contributions from other possible eleven-dimensional geometries are non-negligible. 

It is worth to note some recent advancements in quantum algorithms for lattice gauge theory taking advantage of the large $N$ expansion. 
Ciavarella and Bauer~\cite{Ciavarella:2024fzw} constructed a Hilbert space of gauge invariant states  by acting with single plaquette operators on a vacuum state. 
The leading contributions in large $N$ in the contractible loop sector  comes only from states with isolated single plaquette excitations, and hence the state on each plaquette can be described with the discrete amount of color flux around it.  
With a truncation of the color flux, each plaquette can be mapped to a qubit. In
\cite{Ciavarella:2025bsg}, they demonstrated that the efficient quantum algorithm can also be constructed to subleading order in their formulation.
Their work \cite{Ciavarella:2024fzw} complements our approach~\cite{Berenstein:2023lgo} in the sense that their framework can efficiently describe the states with contractable closed loops, while \cite{Berenstein:2023lgo} is more suitable for the Hilbert space of open strings with fixed quark caps and that of Polyakov loops. 
The open string Hilbert space is separated from the contractable loop Hilbert space as long as quarks have large enough mass, and the Polyakov loop space is separated more strictly due to distinct topological numbers. See also \cite{Modi:2026syn} for other attempts to exploit large $N$ for lattice, in that case with fermions.

In the present article, we review and extend the results of~\cite{Berenstein:2023lgo} to further justify the framework. We explore generalizations with the eventual goal of studying generic confining strings in $3+1$-D (large $N$) QCD. 
The integrability structure of the spin chains describing the first-order perturbation in certain subsectors allows us to analytically calculate the physical quantities of the string, some of which we can use to check if this framework gives a path towards the correct extrapolation to the continuum limit.

The paper is organized as follows. 
In section~\ref{sec:setup}, we review the setting for our framework as the strong-coupling expansion of the Kogut-Susskind Hamiltonian. 
Regarding the magnetic or plaquette terms of the Hamiltonian as perturbation, our framework reduces to a diagrammatic expansion, and its further reduction to the representation as words is justified. 
Section~\ref{sec:1st-order} reviews the integrable spin chains for first-order corrections found in~\cite{Berenstein:2023lgo} and discusses the non-integrability of the more general four letter sector. 
In section~\ref{sec:rot-sym-restoration}, we show the evidence of the validity of this framework by extrapolating the roughening transition point from our perturbation theory calculation, including some subleading corrections. We calculate them in two different string sectors that are expected to coincide in the continuum and demonstrate that they do indeed.
Section~\ref{sec:higher-dimensions} briefly discusses the extension to higher dimensions. 
An example of an integrable spin chain for strings in the $3+1$-D spacetime is shown. 
We also propose another representation of the string words in section~\ref{sec:another-language}, which does not need an explicit projection for the zigzag constraints and hence may be more suitable for some quantum computing applications. 
We conclude this article in section~\ref{sec:discussion} with a discussion of our results and future directions.

\section{Setup \label{sec:setup}}
The Hamiltonian formalism of the lattice pure gauge theory with the $SU(N)$ gauge group and coupling $g$ can be realized with the Kogut-Susskind Hamiltonian \cite{Kogut:1974ag}
\begin{align}
     H_{\text{KS}} &= H_E + H_B
     = \frac{\lambda}{2N}\sum_\ell E_\ell^2 - \frac{N}{2\lambda}\sum_P \Tr[U_P + U_P^\dagger]. 
\end{align}
The first (kinetic) term $H_E$ is a sum over all the lattice links (we call them the electric fields), and the second (potential) term $H_B$ is the sum over all the single plaquettes in the spatial lattice (the magnetic terms). 
Each $E_\ell^a$ ($a = 1,2,...,\deg(SU(N)) = N^2-1$) generates the $SU(N)$ group defined on each link $\ell$ and is in the fundamental representation, and $E_\ell^2 \equiv \sum_a E_\ell^a E_\ell^a$ denotes the quadratic Casimir of $SU(N)$. 
The couplings are written so that $\lambda=g^2N$ is the 't Hooft coupling. The factor $N/\lambda$ of the magnetic term has the standard $N$ dependence associated to large $N$. The kinetic term arises from a Legendre transform of the Kogut-Susskind Lagrangian, which inverts both $\lambda$ and $N$. 
The notion of the electric field is given by 
$E\simeq U^{-1} \dot U$ which acts as a canonical conjugate of $U$ and is actually Lie algebra valued. 
There are actually two possibilities for $E$ defined by the left action of the Lie algebra or the right action. These are related to each other by conjugation by $U$. We choose the left action, so that $U$ is in a fundamental of the left action and $U^\dagger$ is in the antifundamental. The right handed $\tilde E= U E U^{-1}$ is written simply and is useful. Under $\tilde E$, $U$ is in the antifundamental.  We get the same answer as if we are reversing the notion  of $U,U^\dagger=U^{-1}$. 
In order to pick one such choice, we need to add an orientation to each link, which determines in which direction $U$ acts as a matrix.

The quantity $U_P$ is the product of the link operators $U_\ell$ taken along the single plaquette $P$, with $U_\ell$ in the fundamental, so that the orientations are anticyclically ordered. The complex conjugate $U_P^\dagger$ multiplies in the opposite orientation.  
The upper-case trace ($\Tr$ for the second term) is over the fundamental indices to denote that the plaquettes are closed loops, and hence the products have cyclic properties.

The operators defining the Hamiltonian satisfy the following commutation relations : 
\begin{align}
    &[E_\ell^a, E_{\ell'}^b] = if^{abc} E_\ell^c \delta_{\ell\ell'}, \,  
    [E_\ell^a, U_{\ell'}] = T^a U_\ell \delta_{\ell\ell'}, \,  
    [E_\ell^a, U_{\ell'}^\dagger] = -U_{\ell}^\dagger T^a \delta_{\ell\ell'}, 
\end{align}
with the structure constants $f^{abc}$ of $SU(N)$ and the $SU(N)$ generators $T^a$ in the fundamental normalized as $\Tr(T^a T^b) = \delta^{ab}$. Also, all $U$ matrix elements commute with each other at each link and between all links.

The coupling constant $\lambda=g^2N$ is the t' Hooft coupling. 
In the strong-coupling limit $\lambda \gg 1$, the first term dominates and the second term can be regarded as a perturbation with the perturbation parameter $-N/2\lambda$.
The parameter $-N/2\lambda$ is kept small even after we take the large $N$ limit. 
It is contrary to the 't Hooft limit where one takes large $N$ while keeping $\lambda$ constant, so the second term dominates. 
The non-perturbed eigenbasis, i.e. the eigenbasis of the first term has the following property. Since the electric fields commute between different links, each $E^2_\ell$ commutes with the Hamiltonian and all the individual terms in the sum over the links can be diagonalized simultaneously.
To each link we associate a representation of the gauge group 
$G$ and the kinetic term is just the quadratic Cassirmir of said representation. The Hilbert space  is generated as a vector space by
polynomials of the components of $U$ and  $U^{-1}$, with the $1$ function on $G$ acting as the ground state, which has a vanishing Casimir. This ground state $\ket\Omega$ at each link has no energy.
If the representation $R$ is of dimension $\dim R $, the degeneracy of such a basis is actually $\dim R^2$, as the Peter-Weil theorem states that $L^2(G) \simeq \oplus_R R\otimes \bar R$. This is a naive basis where we have not imposed gauge invariance. Since the local Hilbert space is generated 
by the $U$, gauge invariance is implemented at each corner
by contracting $U$ consecutively by matrix multiplication (being mindful of the orientations of the links).

In that sense, gluon excitations are generated by acting with link operators $U$ on it. 

Each $U_\ell$ associated with the link $\ell$ adds a segment starting on one vertex of the link and ending on the other that is dictated by the gauge quantum numbers (and our chosen orientation). 
For example, $U^\dagger_\ell$ adds a link excitation on $\ell$ with the flux in the opposite direction of $U_\ell$, and the unitarity $U_\ell U^\dagger_{\ell'}=\delta_{\ell \ell'}$ indicates that these links cancel each other when we multiply them like a matrix.
These links must form closed directed paths for gauge invariance.
After all the gauge transformations act on each corner by being embedded 
diagonally on the left group actions of $G$ at each link (provided all the orientations are outgoing).
We also consider heavy quarks that allow open strings with fixed endpoints at the location of the quarks. 
Our analysis is in the bulk of the string, but these endpoints are convenient artifacts, so as not to have to deal with the cyclic property of the trace and zero modes. It also allows certain subsectors to be easily identified.

A string is therefore a directed path from one heavy quark to an anti-quark. A path can cross itself. The large $N$ limit guarantees that these possible interconnections do not generate additional energetic contributions, except at subleading orders in $1/N$, which we ignore. The reason for this is that the quadratic Casimir of a representation of $G$ with $k$ boxes at leading order is proportional to $kN$. This also applies if we consider situations with a few boxes and a few antiboxes \cite{Gross:1993hu}.

The non-perturbed energy eigenvalue of such a string state excited along the directed path $\Gamma$, $\ket{\Gamma} \equiv \mc P\prod_{\ell \in \Gamma} U_\ell \ket{\Omega}$ ($\mc P$ indicates the path-ordered product), can be calculated as 
\begin{align}\label{eq:string-energy}
    K\ket{\Gamma}
    &= \frac{\lambda}{2N} \sum_{\ell \in \Gamma} \sum_{a} T^a T^a \prod_{\ell \in \Gamma} U_\ell \ket{\Omega}
    = \frac{N^2 - 1}{2N^2}\lambda L \ket{\Gamma}, 
\end{align}
where $L$ is the length of the path $\Gamma$ in the lattice scale unit, and we used the requirement $E^a_\ell \ket\Omega = 0$ $\forall \ell$ to have the zero ground state energy.  
The value of the Casimir $\sum_a T^a T^a = C_F I = \frac{N^2 - 1}{N} I$ is determined so that it gives the proper normalization $\Tr(T^a T^b) = \delta^{ab}$ in the fundamental representation. 
Since we are interested in the large $N$ limit, the string energy can be further found to be $\lambda L/2$ where $L$ is the total length of the path. 
It is proportional to its length $L$, and the string tension is identified as $\sigma = \lambda/2$. 
Hence, the theory admits confinement with the string tension $\sigma = \lambda/2$ in the strong-coupling limit. 
Since there are many paths between the beginning and the endpoint that can have the same length, the problem of perturbation theory is a degenerate problem. For each fixed length, we need to 
compute the matrix elements of the perturbation in the basis of paths. 

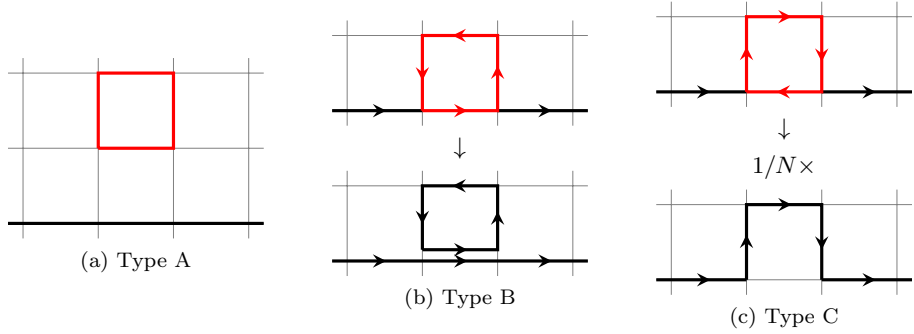
\begin{figure*}[t]
    \centering
    \subfloat[Type A]{\label{subfig:plaq-no-shared}
    \begin{tikzpicture}[baseline = (current bounding box.center)]
        \draw[step=1,gray,very thin] (-1.2,-1.2) grid (2.2,1.2);
        \draw[very thick] (-1.2, -1) -- (2.2, -1); 
        \draw[very thick, red] (0, 0) -- (0, 1) -- (1, 1) -- (1, 0)--(0,0);
    \end{tikzpicture}
    }
    \qquad 
    \subfloat[Type B]{\label{subfig:plaq-same-shared}
    \begin{tikzpicture}[baseline = (current bounding box.center)]
        \draw[step=1,gray,very thin] (-1.2,-1.2) grid (2.2,0.2);
        \draw[very thick, mid arrow=black] (-1.2, -1) -- (0, -1); 
        \draw[very thick, mid arrow=black] (1, -1) -- (2.2, -1); 
        \draw[very thick, red, postaction={on each straight segment={mid arrow = red}}] (0, -1) -- (1, -1) -- (1, 0) -- (0, 0)--(0,-1);
        \node[] at (0.5, -1.5) {$\downarrow$};
        \draw[step=1,gray,very thin] (-1.2,-3.2) grid (2.2,-1.8);
       \draw[very thick, postaction={on each straight segment={mid arrow = black}}] (-1.2, -3) -- (0, -3)-- (1, -3) -- (2.2, -3); 
        \draw[very thick, postaction={on each straight segment={mid arrow = black}}] (0, -2.85) -- (1, -2.85) -- (1, -2) -- (0, -2)--(0,-2.85);
    \end{tikzpicture}
    }
    \qquad 
    \subfloat[Type C]{\label{subfig:plaq-opp-shared}
    \begin{tikzpicture}[baseline = (current bounding box.center)]
        \draw[step=1,gray,very thin] (-1.2,-1.2) grid (2.2,0.2);
        \draw[very thick, mid arrow=black] (-1.2, -1) -- (0, -1); 
        \draw[very thick, mid arrow=black] (1, -1) -- (2.2, -1); 
        \draw[very thick, red, postaction={on each straight segment={mid arrow = red}}] (0, -1) -- (0, 0) -- (1, 0) -- (1, -1)--(0,-1);
        \node[] at (0.5, -1.5) {$\downarrow$};
        \node[] at (0.5, -2) {$ 1/N \times $};
        \draw[step=1,gray,very thin, shift = {(0, 0.5)}] (-1.2,-4.2) grid (2.2,-2.8);
        \draw[very thick, mid arrow=black] (-1.2, -3.5) -- (0, -3.5); 
        \draw[very thick, mid arrow=black] (1, -3.5) -- (2.2, -3.5); 
        \draw[very thick, postaction={on each straight segment={mid arrow = black}}] (0, -3.5) -- (0, -2.5) -- (1, -2.5) -- (1, -3.5);
    \end{tikzpicture}
    }
    \caption{Possible actions of a single plaquette operator on a string state. The cost to attach a plaquette to a string that is already there is $1/N$ ($1/N$ is proportional to the string coupling constant in the t'Hooft counting), but the plaquette is accompanied by $N/\lambda$ so the factors of $N$ cancel. }
    \label{fig:plaq-actions}
\end{figure*}

\begin{figure}
    \centering
    \subfloat[]{\label{subfig:plaq-corner-shared}
    \begin{tikzpicture}[baseline = (current bounding box.center)]
        \draw[step=1,gray,very thin] (-1.2,-1.2) grid (2.2,1.2);
        \draw[very thick, postaction={on each straight segment={mid arrow = black}}] (-1.2, -1) -- (0, -1) -- (1,-1) -- (1, 0) -- (2.2,0); 
        \draw[very thick, red, postaction={on each straight segment={mid arrow = red}}] (0, 0) -- (1, 0) -- (1, 1) -- (0, 1)--(0,0);
    \end{tikzpicture}
    \quad 
    \begin{tikzpicture}[baseline = (current bounding box.center)]
        \draw[step=1,gray,very thin] (-1.2,-1.2) grid (2.2,1.2);
        \draw[very thick, postaction={on each straight segment={mid arrow = black}}] (-1.2, -1) -- (0, -1) -- (1,-1) -- (1, 0) -- (2.2,0); 
        \draw[very thick, red, postaction={on each straight segment={mid arrow = red}}] (0, 0) -- (0, 1) -- (1, 1) -- (1, 0)--(0,0);
    \end{tikzpicture}
    }
    \\
    \subfloat[]{\label{subfig:plaq-corner-shared}
    \begin{tikzpicture}[baseline = (current bounding box.center)]
        \draw[step=1,gray,very thin] (-1.2,-1.2) grid (2.2,1.2);
        \draw[very thick, postaction={on each straight segment={mid arrow = black}}] (-1.2, -1) -- (0, -1) -- (1,-1) -- (1, 0) -- (1,1) -- (0,1) -- (0,0) -- (1,0) -- (2.2,0); 
    \end{tikzpicture}
    \quad 
    \begin{tikzpicture}[baseline = (current bounding box.center)]
        \draw[step=1,gray,very thin] (-1.2,-1.2) grid (2.2,1.2);
        \draw[very thick, postaction={on each straight segment={mid arrow = black}}] (-1.2, -1) -- (0, -1) -- (0.95,-1) -- (0.95, 0) -- (0,0) -- (0,1) -- (1.05,1) -- (1.05,0) -- (2.2,0); 
    \end{tikzpicture}
    }
    \caption{Examples of the type~A' deformations where a string shares a corner with the plaquette (figure~(a)) and gives a nontrivial overlap with another single string state such as states in (b). These contributions are subleading in $1/N$ due to the change of topology. }
    \label{fig:type-Aprime-deformations}
\end{figure}
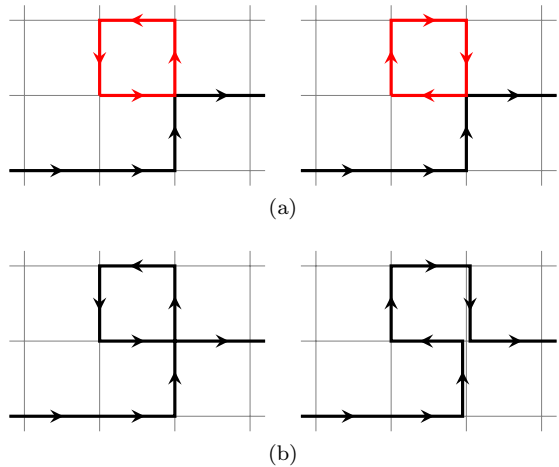

To compute the matrix elements of the perturbation on this basis, one needs to consider the actions of the single plaquette operators on these string states and compute their overlaps with other string states. 
There are three types of actions as discussed in~\cite{Kogut:1974ag} (also see Fig.~\ref{fig:plaq-actions}):
\begin{itemize}
    \item type~A: the plaquette does not share any link with the string. 
    \item type~B: it shares some link(s), which have the same flux direction as the string flux. 
    \item type~C: the shared link(s) have opposite directions. 
\end{itemize}
The details of how to compute overlaps of these deformed string states are described in the next section (section~\ref{sec:string-overlaps}). 
Most of the type~A deformations do not overlap with the open string states and hence do not lead to nontrivial perturbative correction; they are identical to the corrections to the vacuum state. 
There is a special case of the type~A action which nontrivially contributes to corrections higher than the first order. 
It is when the plaquette and the string excitation share one or more vertices. 
Then the state after the plaquette action has an overlap with a single string state with extra energy for an attached loop at the corner (Fig.~\ref{fig:type-Aprime-deformations}). 
We call this type of deformation type~A$'$. 
The overlap is in the order of $\mc O(1/N^c)$, where $c$ is the number of connected corners. 

The type~B deformed state has an overlap in order of $\mc O(1/N^s)$ with a nontrivial string state (namely, a string with $s$ self-intersections on the shared links) that has higher energy than the original string and hence it starts contributing once the second- or higher-order perturbations are considered.

Therefore, we can focus solely on the type~C case for the first-order perturbations. 

For the type~C case, the integral over a single link with opposite flux $\int d U \; {U^i}_j U^{\dagger k}\!_\ell \propto \frac{1}{N} \delta^{i}_\ell\delta^k_j$ indicates that the excitation on the shared link(s) cancels, and hence the plaquette deforms the string with the extra factor $1/N$. 
The integral arises from an overlap calculation between the states at the individual link. After all, this is how we write the inner product in $L^2(G)$ of the corresponding link in the position basis.
The cancellation allows the deformed string to possibly have the same energy as the original string, so it may contribute to the first-order degenerate perturbations. 
This extra $1/N$ factor cancels the usual $N$ factor in $H_B$.
Therefore, it results in a finite contribution to the energy.
The large-$N$ limit also ensures the locality of the effective theory on the string, similar to how it works in the ${\cal N}= 4$ SYM spin chain \cite{Berenstein:2002jq} (that is, we ignore self-intersections of the string with itself when discussing corrections: these would be non-local on the string, and they are suppressed by extra factors of $1/N$).

We label a string state by the path it takes. 
The string is a word with instructions on what the next step is: up ($u$), down ($d$), left ($l$), right ($r$), etc. 
Each of these is treated as a letter of the alphabet of paths. 
An action of the plaquette changes the shape of the path. 
The $UU^\dagger=1$ constraint is implemented by stating that a path cannot double back on itself immediately. 
That is, combinations such as  $\dots ud\dots$, or $\dots l r\dots$ are forbidden. 
This is also called zigzag symmetry and is related to the reparametrization invariance on the worldsheet~\cite{Polyakov:1997tj}.
A way to think about this is that in this case (the string in $2+1$ dimensions), if at some point a letter is $u$, then after it there are only three possibilities $u,l,r$, so the number of states at length $L$ is $4\times 3^{L-1}$, as the first letter leaving the quark is unconstrained.
If we insist on using the 4 letters, the Hilbert space is the set of words with $ud$, $du$, $lr$ and $rl$ combinations forbidden: we can write a projection operator $\pi$ that projects onto the allowed space of letters at each link. It acts by $\pi(lr)=0$, etc,
on the forbidden combinations
but $\pi(dl)= dl$ and similar for the allowed configurations. Since these are orthogonal to each other $\pi$ is self-adjoint.

\subsection{Norms and overlaps of the string states\label{sec:string-overlaps}}
The calculations of the overlaps between states with stringy excitations are achieved by path-integrating the link variables over the gauge group manifold in the general form of 
$\int \left(\prod_\ell dU_\ell\right)\, f(\{U_\ell\})$.
Such integrals over the paths can be packaged into diagrammatical computations in our Hamiltonian formalism with some rules. 
Here, we list the diagrammatical rules for computing overlaps between stringy states and use them to demonstrate that the first-order corrections to single strings are indeed closed in large~$N$, and hence their representations as words are valid. 

The rules we must apply include the requirement of specifying the directions of the graph edges, the proper normalization of the nonperturbed vacuum states, contractions due to the zigzag (or unitarity) constraints, and the normalization of the trace over the fundamental indices. 
\begin{enumerate}
    \item {\textbf{Directed edges:} whether a link excitation corresponds to the operator $U$ or $U^\dagger$ is specified with a directed edge, with the endpoints specified with its fundamental indices. On the same link, $U$ and $U^\dagger$ have opposite directions. \label{rule:graph}}
    \item {\textbf{Vacuum normalization:} the nonperturbed vacuum state is properly normalized: $\braket{\Omega}{\Omega} = 1$. This is also automatically true on any link without lines going through it. \label{rule:vacuum-norm} }
    \item {\textbf{Contractions:} for the edges on the same link, from the group integral computation, the following contraction occurs due to the unitarity of the link operators: 
\begin{align}
        U_{ij}U_{kl}^\dagger 
        &= 
        \vcenter{\hbox{\includegraphics{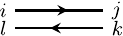}}}
        =  \frac{1}{N} \times \left(
        \vcenter{\hbox{\includegraphics{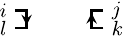}}}
        \right) 
        = \frac{1}{N}\delta_{il}\delta_{jk}. 
\end{align}
By contracting the $j$ and $k$ indices, we obtain the unitarity constraint:
\begin{align}
        U_{ij}U_{jk}^\dagger 
        &= 
        \vcenter{\hbox{\includegraphics{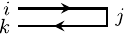}}}
        =  
        \vcenter{\hbox{\includegraphics{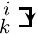}}}
        = \delta_{ik}. 
    \end{align}
    Contractions involving higher power of the $U, U^\dagger$ operators can also be computed using the group integrals. 
    Our discussion will only involve the second-power: 
    \begin{widetext}
    \begin{align}\label{eq:U-Udag-U-Udag}
        U_{ij}U^\dagger_{kl}U_{mn}U^\dagger_{pq}
        &= 
        \vcenter{\hbox{\includegraphics[]{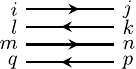}}}
        =
        \frac{1}{N^2 - 1}\bigg(
        \vcenter{\hbox{\includegraphics[]{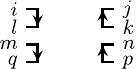}}}
        +
        \vcenter{\hbox{\includegraphics[]{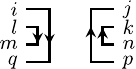}}}
        + \mc O\left(\frac{1}{N}\bigg)
        \right)\nn 
        &\sim 
        \frac{1}{N^2}(\delta_{il}\delta_{jk}\delta_{mq}\delta_{np}
        + \delta_{iq}\delta_{jp}\delta_{lm}\delta_{kn}),
    \end{align}
    \end{widetext}
    where $\sim$ means the contribution in the large $N$ limit. 
    \label{rule:contraction}}
    \item {\textbf{Fundamental trace:} a zero-length closed loop, corresponding to $\Tr(I)$, contributes by a factor of $N$. \label{rule:zero-loop}}
\end{enumerate}
First, notice that these requirements guarantee that all of the nonperturbed eigenstates are normalized properly. 
For example, let us compute the norm of a single open string state diagrammatically.
The ket and bra states correspond to a path going forward and backward, respectively, and they connect with each other at one of the endpoints once we take the inner product. 
The folded string fully contracts and ends up with an identity operator: 
\begin{align}
    \braket{\Gamma}{\Gamma}
    &= 
    \bra\Omega 
    \vcenter{\hbox{\includegraphics{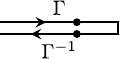}}}
    \ket\Omega
    = 
        \bra\Omega
        \vcenter{\hbox{\includegraphics{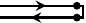}}}
        \ket\Omega
        = \cdots 
        = 
        \bra\Omega \tikz [] {
        \draw[very thick] (0, 0) -- (0.1, 0) -- (0.1, -0.2) -- (0, -0.2); 
        }\ket\Omega
        = \braket{\Omega}{\Omega}
        = 1
\end{align}
For the other open string states, we can apply a similar backtracking argument. 
The other nontrivial cases are the states containing closed-loop excitations. 
Let us confirm that these states also are normalized as well by considering a single loop as an example: 
\begin{align}
    \braket{C}{C}
    &= 
    \bra\Omega 
    \vcenter{\hbox{\includegraphics[]{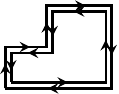}}}
    \ket\Omega 
    = 
    \frac{1}{N^6} \times \bra\Omega 
    \vcenter{\hbox{\includegraphics[]{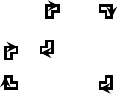}}}
    \ket\Omega
    = 1
\end{align}
Notice that closed loops have the same number of vertices and edges, so the $1/N$ factors from the edge contractions are canceled exactly with the zero-length closed loops remaining at the vertices following rule~\ref{rule:zero-loop}. 

We can easily show the orthogonality of the states with excitations on different edges. 
In general, the nonperturbed vacuum expectation values of gluon link excitations on a path not contractable by rule~\ref{rule:contraction} to zero length are required to be zero. 
This can be easily proven by picking one of the remaining gluon operators $U_\ell$ after all possible contractions and inserting the $E^a E^a$ operator acting on the link $\ell$, which automatically gives zero acting on the vacuum state: 
\begin{align}
    0 &= \mel{\Omega}{\cdots U_\ell \cdots E^a_\ell E^a_\ell}{\Omega}
    = \mel{\Omega}{T^a T^a \cdots U_\ell \cdots}{\Omega}
    = \frac{N^2 - 1}{N}\mel{\Omega}{\cdots U_\ell \cdots}{\Omega}
\end{align}
for the single $U_\ell$ case (i.e. other $U$ operators are acting on different links). 
A similar but more involved arguments apply to prove this statement for the states involving higher powers of $U_\ell$. 

\begin{figure}[t]
    \centering
    \subfloat[]{\label{fig:string-twisted}
    \includegraphics[]{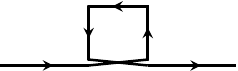}
    }
    \quad 
    \subfloat[]{\label{fig:string-plus-loop}
    \includegraphics[]{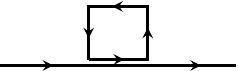}
    } 
    \caption{An example of string configurations with excitations on the same links but with different connectivity. }
    \label{fig:string-one-loop}
\end{figure}

The remaining nontrivial overlaps are the ones between two string states with the gluon excitations on the exactly same links but with different connectivity. 
These overlaps are all subleading in $1/N$, since the processes of changing the connectivity from one to another are nonplanar. 
Specifically, each crossing or overlapping graph can be exchanged to the other with a factor of $1/N$: 
\begin{align} \label{eq:cross-overlap}
    \begin{tikzpicture}
        \draw[very thick, -stealth] (0, 0) -- (2, 0);
        \draw[very thick, -stealth] (0, -0.2) -- (2, -0.2);
    \end{tikzpicture}\; 
    \xleftrightarrow{\times 1/N}\; 
    \begin{tikzpicture}
        \draw[very thick, -stealth] (0, 0) -- (2, -0.2);
        \draw[very thick, -stealth] (0, -0.2) -- (2, 0);
    \end{tikzpicture}
\end{align}
and hence the inner products scale as $\sim 1/N^s$ where $s \in \mZ_{\geq 0}$ is the number of different crossings/overlaps as Eq.~\eqref{eq:cross-overlap}. 
Let us diagrammatically calculate an example overlap between an open string with an attached twist (Fig.~\ref{fig:string-twisted}), which can be represented as a single word, and a string with a separate loop (Fig.~\ref{fig:string-plus-loop}), which cannot be represented as only one word. 
They have the same excited links, but different connectivity only around the twist or the loop. 
The overlap of these two states is at the next leading order of $ 1/N$: 
\begin{widetext}
\begin{align}\label{eq:overlap-twist-loop}
    \braket{
    \vcenter{\hbox{\includegraphics[]{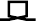}}}
    }
    {
    \vcenter{\hbox{\includegraphics[]{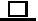}}}
    }
    &= 
    \bra \Omega
    \vcenter{\hbox{\includegraphics[]{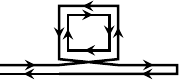}}}
    \ket \Omega\nn 
    &= 
    \frac{1}{N^2 - 1}\bra \Omega \left(
    \vcenter{\hbox{\includegraphics[]{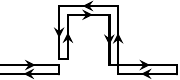}}}
     + 
     \vcenter{\hbox{\includegraphics[]{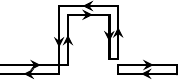}}}
\right)
    \ket \Omega\nn 
    &
    \sim \frac{2}{N}, 
\end{align}
\end{widetext}
where the second equation is obtained by the contraction of $UU^\dagger UU^\dagger$ (Eq.~\eqref{eq:U-Udag-U-Udag}). 
The combinatorial factor of two appears due to the two ways of contraction contributing by an equal amount. 
For the first-order corrections in the large $N$ limit, we can ignore these overlaps between states with different topologies, which means that the first-order corrections are closed in the sector of single string states with equal length. 
Therefore, it is justified to represent the single strings as words of letters specifying the directions and to reduce the plaquette actions as manipulations of the letters. 
On the other hand, since we take the double scaling limit $\lambda \gg N \gg 1$, these nonplanar contributions may contribute at higher-orders in perturbation theory. 
Note that the discussion in this section is fully spacetime dimension independent, even though we carried out the above example calculations in $2+1$-D. For the general case, we just have more letters (two for each spacetime direction and a similar restriction on consecutive letters).

\section{First-order corrections\label{sec:1st-order}}
\begin{figure}[t]
    \centering
    \includegraphics[width=\linewidth]
    {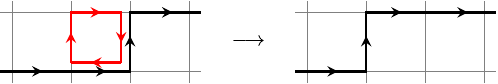}
    \caption{One example of the type~C deformations by a single plaquette action (red in the left figure) generating another degenerate state. 
    Notice that the deformed string has exactly the same string length as the original state. 
    This plaquette action exchanges the $u$ and $r$ letters next to each other. }
    \label{fig:typeC-degenerate}
\end{figure}

Here, we review and expand the results of~\cite{Berenstein:2023lgo} in $2+1$-D with some generalizations and introduce the concept of ``knottiness", which is a conserved quantity in a string sector in $2+1$ dimensions due to the zigzag constraint. 
As discussed in Sec.~\ref{sec:setup}, only type~C deformations in Fig.~\ref{subfig:plaq-opp-shared} give contributions to the first-order corrections, and they can be described as manipulations of letters as one describes the single string states as words of letters. 
The letters can be considered as quantum spin states, and the letter manipulations as the actions of spin operators. 
The general string states in $2+1$-D are described with words containing four letters: $u, d, l, r$. 
The type~C deformations act as the exchange of nearby $u, d\leftrightarrow l,r$ and similar. That is, there needs to be a hook at the location of the plaquette, so that two links are destroyed and two other links are created (Fig.~\ref{fig:typeC-degenerate}).
These must be going in perpendicular directions.

The extra factor $1/N$ due to the deformation cancels $N$ in the numerator of the coupling factor of $H_B$, namely $-N/2\lambda$.  
The first-order corrections are closed under such degenerate type~C deformations within the sectors of the fixed number of each letter. 
That is, the effective spin operator preserves the number of $u,d,l,r$ independently of each other.

The first-order corrections are obtained by diagonalizing the matrix with the elements of $ \mel{\Gamma_i}{\Pi \Tilde H_B \Pi}{\Gamma_j}$, where $\ket{\Gamma_i}$ are the single string states, and $\Pi$ is the projection operator prohibiting $u$ and $d$ or $r$ and $l$ to be next to each other. This needs only be applied to the final state if the initial state is already allowed by the projector.
$\Tilde H_B$ is the sum of the letter manipulations on a word as 
\begin{align}
   \Tilde H_B &= \sum_\ell \Tilde h_\ell \nn 
   &= -\frac{1}{2\lambda} \sum_\ell (u\text{ or } d \text{ at }\ell \leftrightarrow r \text{ or } l \text{ at }(\ell + 1)) 
   +  (r\text{ or } l \text{ at }\ell \leftrightarrow u \text{ or } d \text{ at }(\ell + 1)), 
\end{align}
where $\ell$ runs through every letter position of the word specifying the string configuration. 

A similar way to write this is
\begin{align}
\tilde H_B &= \sum_\ell \ket{ur}\bra{ru}+\ket{dr}\bra{rd}
+
\ket{ul}\bra{lu}+\ket{dl}\bra{ld}+c.c. 
\end{align}

And now we need to introduce the projector onto allowed states 
where $\Pi= \prod_\ell \pi_\ell$.
The full first-order perturbation operator is $\Pi \Tilde H_B \Pi$, which has a local description 
\begin{align}
    \Pi \Tilde H_B \Pi = \sum_\ell (\pi_{\ell-1}\otimes \pi_{\ell+1}) \Tilde h_\ell (\pi_{\ell-1}\otimes \pi_{\ell+1}),
\end{align}
and if the initial states are eigenstates of $\Pi\ket \Gamma=\ket{\Gamma}$, 
then the rightmost $\Pi$ can be ignored. 

Basically, here the $\pi_\ell$ operator projects out the state if its local configuration at $(\ell - 1, \ell)$ or $(\ell+1, \ell+2)$ has suddenly become forbidden $u$ and $d$ or $r$ and $l$ next to each other. 
This means that  the projected operator $\Pi \Tilde H_B \Pi$ consists of local operators acting on four neighboring sites, rather than two, but the basic operator $\Tilde H_B $ is only a nearest neighbor process.
Indeed, if we ignore the $\Pi$ altogether, which means we ignore the zigzag constraints,  the nearest neighbor  Hamiltonian described exactly by $\tilde H_B$ is integrable, a result we found in a previous paper \cite{Berenstein:2023lgo}. What this means is that the most important part of the analysis to test integrability or lack thereof needs to take the constraints imposed by $\Pi$, or equivalently the different $\pi$ seriously.

We start our discussion with some simple sectors: the sectors of the configurations containing only two letters, namely $u$ and $r$, and the sectors with only three letters, say $u, d, r$. 
We shall demonstrate that these two kinds of closed subsectors are integrable by translating this operator into spin chains only with nearest-neighbor interactions, whereas the general four-letter configuration sectors cannot. 
Specifying the two-letter or three-letter sector with the number of each letter corresponds to specifying the fixed end points or heavy nondynamical quarks. 
As the theory goes to continuum by having the lattice spacing $a \rightarrow 0$, the rotational symmetry must be recovered, and hence the low energy spectra of these two configurations must coincide. 
Once we promote the Yang-Mills action to QCD by turning on the fermion dynamics, the end points would be no longer fixed and these sectors would start mixing. 
We will leave this to future work. Even just pure glue with heavy quark endpoints is interesting on its own.

Let us comment on the projection by $\Pi$. 
The projection prohibits changing the  orderings of $u, d$ with each other and $l, r$ with each other. 
We call these protected orderings the ``knottiness". 
In a word with $4\times 3^{L-1}$ states these would give $2^{L}$ superselection sectors that cannot mix with each other.
A similar result would occur in higher dimensions. For the $2+1$-D string there are additional conservation laws, which we will call ``secondary knottiness" and we will  explain it when we consider the full problem with four letters.

\begin{figure}
    \centering
    \subfloat[]{\label{fig:orientation-original} 
    \includegraphics[width=0.45\linewidth]{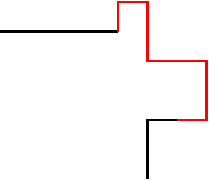}
    }
    \, 
    \subfloat[]{\label{fig:orientation-ordered} 
    \includegraphics[width=0.45\linewidth]{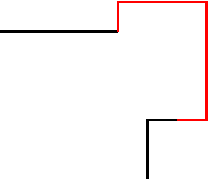}
    }
    \caption{An example of evaluating the local orientation of the string. 
    (a) Let us focus on the substring $rrrr\textcolor{red}{urddrrddl}ldd$ (depicted in red). It is bounded by $u$ and $l$. 
    (b) Exchanging the positions of the bulk letters $r$ and $d$, the substring is now $\textcolor{red}{urrrddddl}$. The orientation is determined to be clockwise from this procedure. }
    \label{fig:orientation}
\end{figure}

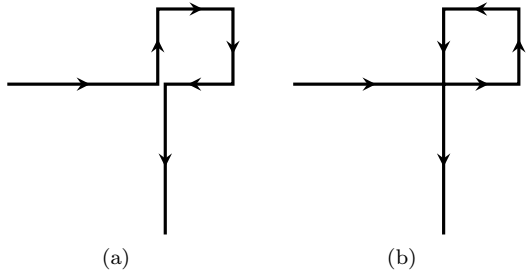
\begin{figure}
    \centering
    \subfloat[]{\label{fig:corner-cw} \begin{tikzpicture}
        \draw[very thick] (0, -0.5) -- (2, -0.5) -- (2, 0.5) -- (3, 0.5) -- (3, -0.5) --(2.1, -0.5) -- (2.1, -2.5); 
        \draw[-stealth, very thick](1, -0.5)--(1.1, -0.5); 
        \draw[-stealth, very thick](2.5, -0.5)--(2.4, -0.5); 
        \draw[-stealth, very thick](3, 0)--(3, -0.1); 
        \draw[-stealth, very thick](2.5, 0.5)--(2.6, 0.5); 
        \draw[-stealth, very thick](2, 0)--(2, 0.1); 
        \draw[-stealth, very thick](2.1, -1.5)--(2.1, -1.6); 
    \end{tikzpicture}}
    \qquad 
    \subfloat[]{\label{fig:corner-ccw} \begin{tikzpicture}
        \draw[very thick] (0, -0.5) -- (3, -0.5) -- (3, 0.5) -- (2, 0.5) -- (2, -2.5); 
        \draw[-stealth, very thick](1, -0.5)--(1.1, -0.5); 
        \draw[-stealth, very thick](2.5, -0.5)--(2.6, -0.5); 
        \draw[-stealth, very thick](3, 0)--(3, 0.1); 
        \draw[-stealth, very thick](2.5, 0.5)--(2.4, 0.5); 
        \draw[-stealth, very thick](2, 0)--(2, -0.1); 
        \draw[-stealth, very thick](2, -1.5)--(2, -1.6); 
    \end{tikzpicture}}
    \caption{An example of two string states excited on the same edges but with different local orientations. The zigzag constraint prohibits any Type~C deformations to deform one to the other, and their overlaps are $1/N$-suppressed due to the different connectivity. These differ by what we termed secondary knottiness.}
    \label{fig:corner}
\end{figure}

Additionally, the $U(N)$ orientations leads to a protection of the orientation of the string: it distinguishes strings going from quark one to quark two from the opposite direction.
The orientation can be evaluated as follows (see also Fig.~\ref{fig:orientation} for an example). 
We choose some substring of the word containing all four kinds of letters but with the minimum length.
This minimal substring contains a pair of two perpendicular letters on the left and right ends, and the bulk is filled with the arbitrary length of the other two letters. 
One can act with Type~C deformations to exchange the positions of the bulk letters, so that the same kind of letters are localized together. 
Now, the four kinds of the letters are ordered either (i) clockwise, i.e. ``$urdl$" or its cyclic permutations, or (ii) counter-clockwise, i.e. ``$uldr$" or its cyclic permutations. 
These two orientations cannot be reached from one to the other for any subset of the word containing four kinds of letters by any degenerate action of plaquette operators. 
It would require an exchange of either $u \leftrightarrow d$ or $r \leftrightarrow l$. 
Moreover, although it is possible to consider string excitations on the same edges with different local orientations, their overlaps are suppressed by at least $1/N$ (Fig.~\ref{fig:corner})

The way to describe what configurations are related to each other is that we need to keep track of zigs and zags (horizontal and vertical changes of directions). Call these $h,v$. A combination $urd$ would count as an $h$ as there is a vertical orientation shift, but $uru$ would not. Similarly $rul$ would count as $v$, but not $rur$.
The pattern of $hv$ is also assembled into an ordered string as we move along the flux tube.
This pattern of $hv$ words cannot be modified either by the plaquette actions. Basically, it would need to change $urdl$ to the opposite order of $hv$, $ruld$, but such a move is not allowed.
What this means is that although three letter superselection sectors are relatively easy to analyze (there are only $v$ swicthbacks let's say), the four letter ones require this extra superselection label and become more complicated.

\subsection{Two-letter strings\label{sec:two-letter-1st}}
As we mentioned above, the action of $\Tilde H_B$ is closed within the subsectors of strings with a fixed combination of the letter numbers. 
The simplest nontrivial subsector we can consider contains only the strings with two letters. Strings with one letter are trivial: the
Hamiltonian does not act on them keeping the length fixed, so there is nothing to calculate in the first order.

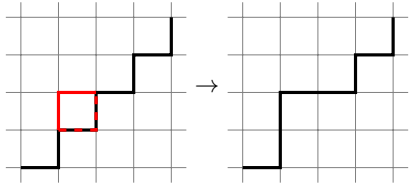
\begin{figure}[t]
    \centering
    \begin{tikzpicture}[baseline = (current bounding box.center)]
        \draw[step=0.5,gray,very thin] (-1.2,-1.2) grid (1.2,1.2);
        \draw[very thick] (-1, -1) -- (-0.5, -1)--(-0.5, -0.5)--(0, -0.5)--(0, 0) --(0.5, 0) -- (0.5, 0.5)--(1, 0.5)--(1, 1); 
        \draw[very thick, red] (-0.5, -0.5) -- (-0.5, 0) -- (0, 0);
        \draw[very thick, red, dashed] (-0.5, -0.5) -- (0, -0.5) -- (0, 0);
    \end{tikzpicture}
    $\large \rightarrow$
    \begin{tikzpicture}[baseline = (current bounding box.center)]
        \draw[step=0.5,gray,very thin] (-1.2,-1.2) grid (1.2,1.2);
        \draw[very thick] (-1, -1) -- (-0.5, -1)--(-0.5, -0.5)--(-0.5, 0)--(0, 0) --(0.5, 0) -- (0.5, 0.5)--(1, 0.5)--(1, 1); 
    \end{tikzpicture}
    \caption{The diagonal string $\Gamma$ with $m$ vertical links and $L-m$ vertical links (thick black path).
    If one acts the single plaquette operator (red) sharing two links (red dotted lines) on the diagonal string state, it changes the state to another with the same string length and the endpoints. }
    \label{fig:diagonal-plaquette}
\end{figure}

This subsector of two letters is the simplest in the sense that we do not need the projection $\Pi$ as all possible combinations are allowed. 
Let us choose the subsector containing $n$ of $u$ and $L-n$ of $r$. 
One possible excitation is along the path as in the left figure of Fig.~\ref{fig:diagonal-plaquette} with length $L$. 
The endpoint quarks are placed diagonally from the lattice axes in general. 
The single plaquette operator sharing exactly two links with the original configuration $\Gamma$ as the red plaquette in Fig.~\ref{fig:diagonal-plaquette} acts a type~C deformation and generates another string excitation generated with the same string length and the same positions of the quark-antiquark endpoints.
Hence, the states degenerate with the excitation along $\Gamma$ are generated by a series of this kind of plaquette moves. 
The Hilbert space, which we call $\mc H_{\Gamma}$, spanned with these degenerate states has a dimension of ${ L \choose n}$. 

The first-order corrections to the strong-coupling limit eigenstates can be found by diagonalizing the ${ L \choose n}\times { L \choose n}$ matrix 
\begin{align}
    W_{ij} \equiv \mel{\Gamma_i}{\Tilde H_B}{\Gamma_j}
\end{align}
where $\ket{\Gamma_i} \in \mc H_\Gamma$.
The action of a single plaquette operator sharing two of its edges with the string can be seen as flipping a neighborhood vertical link $u$ and a horizontal link $r$. 
We translate them to the spin-$1/2$ states, $u \leftrightarrow \uparrow$ and $r \leftrightarrow \downarrow$. 
In this language, the plaquette operator flips the spins of two neighborhood spin-$1/2$ sites with opposite spins, i.e. 
\begin{align}
    \Tr[U_P+U_P^\dagger] \leftrightarrow \frac{1}{N}(\sigma^+_ j\sigma^-_{j+1} + \sigma^-_j \sigma^+_{j+1})
\end{align}
where the spin operators 
act on each spin-$1/2$ state as $\sigma^+ \ket \downarrow = \ket \uparrow$, $\sigma^- \ket \uparrow = \ket \downarrow$, and $\sigma^+ \ket \uparrow = \sigma^- \ket \downarrow = 0$. 
The whole matrix $W_{ij}$ coincide with the matrix representation of this spin operator summed over all the spin sites. 
Hence, the action of the plaquette perturbation $V$ ends up being exactly the XX model Hamiltonian  
\begin{align}\label{eq:H-XX}
     H = -\frac{1}{2\lambda} \sum_{j = 1}^{L} (\sigma^+_ j\sigma^-_{j+1} + \sigma^-_j \sigma^+_{j+1}). 
\end{align}
This Hamiltonian conserves the magnetization, and one can decompose the Hilbert space into the sectors with different magnetization values as follows
$
     H_{\text{XX}} = \bigoplus_{m = 0}^{L} \mc H_m
$
where $\mc H_m$ is the sector having $m$ down spins, whose dimension is ${ L \choose m} $. 
The different subsectors exactly correspond to the different combinations $(n, L-n)$ for the numbers of $u$ and $r$. 
The first-order perturbation matrix $W_{ij}$ is the block of the XX model Hamiltonian in $\mc H_{n}$. 
The XX model is trivially integrable given that it is equivalent to two copies of a free massless fermion theory. It is also a critical theory.
We take this to mean that the effective sigma model on the string world sheet is that of a $c=1$ theory. It is two free massless fermions, but it can be bosonized to a $c=1$ free boson, which is interpreted as the goldstone boson of translations that are spontaneously broken by the string. These bosons should be thought of as fluctuations about the diagonal. 

Kogut et al. found this free fermion description for the diagonal string setup in~\cite{Kogut1981} and studied just the ground state of the string
in order to find how rotational symmetry is restored when the coupling becomes finite.
Nearest-neighbor spin chains such as these are easily implementable in quantum computers, so it could be interesting to study states that are dynamical and not only the ground state.

The fact that this subsector is integrable should be no surprise: spin chains with only nearest-neighbor interactions preserving the spin $\sigma_z$ and parity conserving (treat $\uparrow\downarrow$ equivalently) are all members of the XXZ family.

\subsection{Three-letter strings\label{sec:three-letter-1st}}
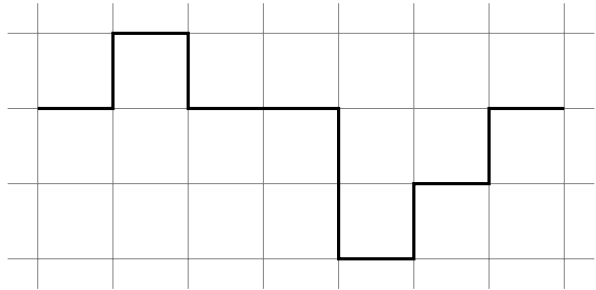
\begin{figure}[t]
    \centering
     \begin{tikzpicture}[scale = 2, baseline = (current bounding box.center)]
        \draw[step=0.5,gray,very thin] (-1.7,-1.2) grid (2.2,0.7);
        \draw[very thick] (-1.5, 0) -- (-1, 0)--(-1, 0.5)--(-0.5, 0.5) -- (-0.5, 0) -- (0, 0) -- (0.5, 0) -- (0.5, -1) -- (1, -1) -- (1, -0.5) -- (1.5, -0.5) -- (1.5, 0) -- (2, 0);
    \end{tikzpicture}
    \caption{An example of three-letter configuration. It corresponds to the word $rurdrrddrurur$. \label{fig:three-letter-string}}
\end{figure}

We now turn to the second case where string excitations consist of three letters.
Let us choose $u, d, r$, that is we allow the link excitations to be up, down, or right.
General three-letter configurations look like Fig.~\ref{fig:three-letter-string}. 
To characterize these string configurations, we express them with three letters $u$, $d$, and $r$ respectively, being careful to disallow $ud$ and $du$ words because we are implementing the zigzag symmetry. 
These words are identical to the configurations with a shorter string length and do not reside in the degenerate sector. 
The energy-conserving action of a plaquette on $ur$ and $dr$ pieces of the string looks the same as that for the diagonal string case, which allows swapping the letters as $u(d)\leftrightarrow r$. 
The new wrinkle is that swapping the order $urd$, for example, can land one in the forbidden configuration $ud$.
In that sense, those possibilities must be projected out. 
Such projections effectively turn the problem into one where the swap of letters is allowed depending on extra letters, a three- or four-letter problem. The upshot is that at this order in perturbation theory, $ud$ letters can not swap their position. 
This property is similar to that of the eclectic and hyper-eclectic spin chain where some letters (states) act sometimes as walls \cite{Ipsen:2018fmu,Ahn:2020zly}.

It is convenient to pass to a new basis, where we treat $ud$ as walls, and use an occupation number basis for $r$ in between each pair of walls. This is identical to the treatment employed to describe the XXX spin-chain with variable length in~\cite{Berenstein:2005fa} (see also \cite{Berenstein:2006qk}). 
In this basis, the Hamiltonian for the perturbation becomes
\begin{equation}
   H=- \frac{1}{2\lambda} \sum (c_i^\dagger c_{i+1} +c_i c^\dagger_{i+1})
\end{equation}
where $c_i$ are the Cuntz oscillators satisfying $c_i c_i^\dagger=1$, and they commute between different sites $i\neq j$.
Their action is $c^\dagger \ket n= \ket{n+1}$ and $c\ket 0=0$ for each Fock space. 
We restrict the Hilbert space by requiring that the occupation number between any walls that change direction $ud$ or $du$ is greater than or equal to one. 
This means that we do not allow the $ud$ to be neighbors without any restrictions. 
It turns out that this restriction leaves the Cuntz algebra unchanged on the relevant states as if the occupation one state had occupation number zero: the algebra does not change if we do a shift by one plus a projection.  In that sense, we can relabel the states with a modified Cuntz oscillator at these sites that ignores that constraint: it becomes an implicit padding of the words whenever it is required. That is, we take advantage of the basis change:   
\begin{align}\label{eq:3-letter-padding}
    &\ket 1 \rightarrow \ket 0: dru \rightarrow du, \nn 
    &\ket 2 \rightarrow \ket 1: drru\rightarrow dru, \nn 
    & \hspace{2em} \vdots 
\end{align}
This padding now allows us to describe the whole problem as a nearest neighbor spin chain without extra projections \cite{Berenstein:2023lgo}. 

The result is a simple Hamiltonian where one only swaps $d \leftrightarrow r$ and $u\leftrightarrow r$ and their reverse swaps are allowed. 
All other terms in the Hamiltonian vanish. We can map these to a spin-one system as $r\to\ket 0$, $u\to \ket +$, $d\to \ket -$. 
The Hamiltonian is equivalent to the following spin-chain Hamiltonian acting on this Hilbert space $\mc H = (\mC^3)^{\otimes L}$:  
\begin{align}\label{eq:H-straight}
     &H = 
     -\frac{1}{2\lambda}  \sum_{j = 1}^L \left(e^1_j f^1_{j+1} + f^1_j e^1_{j+1} + e^2_j f^2_{j+1} + f^2_j e^2_{j+1} \right)
\end{align}
with the generators of $\mathfrak{sl}(3, \mC)$:
\begin{widetext}
\begin{align}\label{eq:gen-sl3}
    h^1 &= 
    \begin{pmatrix}
        1 & 0 & 0\\ 
        0 & -1 & 0\\
        0 & 0 & 0
    \end{pmatrix}, 
    \; 
    h^3 = 
    \begin{pmatrix}
        0 & 0 & 0\\ 
        0 & 1 & 0\\
        0 & 0 & -1
    \end{pmatrix}
    \nn
    e^1 &= 
    \begin{pmatrix}
        0 & 1 & 0\\ 
        0 & 0 & 0\\
        0 & 0 & 0
    \end{pmatrix}, 
    \; 
    e^2 = 
    \begin{pmatrix}
        0 & 0 & 0\\ 
        0 & 0 & 0\\
        1 & 0 & 0
    \end{pmatrix}
    , \; 
    e^3 = 
    \begin{pmatrix}
        0 & 0 & 0\\ 
        0 & 0 & 1\\
        0 & 0 & 0
    \end{pmatrix}
    \nn
    f^1 &= 
    \begin{pmatrix}
        0 & 0 & 0\\ 
        1 & 0 & 0\\
        0 & 0 & 0
    \end{pmatrix}, 
    \; 
    f^2 = 
    \begin{pmatrix}
        0 & 0 & 1\\ 
        0 & 0 & 0\\
        0 & 0 & 0
    \end{pmatrix}, 
    \; 
    f^3 = 
    \begin{pmatrix}
        0 & 0 & 0\\ 
        0 & 0 & 0\\
        0 & 1 & 0
    \end{pmatrix}
\end{align}
\end{widetext}
$L$ is the length of the words. 
The subscript $j$ specifies that the operator is acting on the $j$-th spin site. 
Each term exchanges the neighbor $\ket 0$ and either $\ket{+}$ or $\ket{-}$. 
$\ket{+}$ and $\ket{-}$ cannot pass through each other, as the exchange $u\leftrightarrow d$ is forbidden.
A similar model was found independently by \cite{Alcaraz:1991ps} and \cite{Gomez:1992nd} with the only difference being  the additional chemical potential term $\propto i$ \footnote{The authors would like to thank Hosho Katsura for pointing this out. }, where $ i = \sum_{j=1}^L i^3_j$ with 
\begin{align}
    i^3 = \begin{pmatrix}
        0 & 0 & 0\\
        0 & 1 & 0\\
        0 & 0 & 1
    \end{pmatrix}. 
\end{align}
This model has an $SU(2) \times U(1)_i$ symmetry generated by $ h = \sum_{j=1}^L h^3_j$, $ e = \sum_{j=1}^L e^3_j$, and $ f = \sum_{j=1}^L f^3_j$ for the $SU(2)$, and $i$ for $U(1)_i$. 
One can interpret the $SU(2)$ part as the invariance of the system under the exchange of $\ket{+}$ $\leftrightarrow $ $\ket{-}$. 
For each spin site, the $\ket{0}$ state forms a singlet under this $SU(2)$, and $\ket{\pm}$ forms a doublet. 
That is, each spin site is in the representation of $\mathbf 1 \oplus \mathbf 2 = \mathbf 1 \oplus \ydiagram{1}$ of this $SU(2)$ symmetry. 
The Hilbert space of the whole spin chain can be decomposed as
\begin{align}\label{eq:su2-rep}
    \mc H 
    &= \left(\mathbf 1 \oplus \ydiagram{1}\right)^{\otimes L}
    = 
    \mathbf 1 \oplus 
    \ydiagram{1}^{L}
    \oplus 
    \ydiagram{2}^{{L\choose 2}}
    \oplus 
    \ydiagram{1,1}^{{L\choose 2}}
    \oplus \cdots 
\end{align}
The number of the Young boxes corresponds to the $U(1)_i$ charge of the state in the representation, i.e. the total number of $\ket +$ and $\ket -$. 
The state in the first singlet is $\ket \Omega = \ket{00...0}$, which we choose as the Bethe reference state for solving this model in Appendix~\ref{sec:three-letter-Bethe-ansatz}. 
The theory has only elastic scatterings due to the conservation of the two $U(1)$ charges, corresponding to the $U(1)$ sub-symmetries generated each by $i$ and $h$ operators defined above. 
The conservation ensures that the $\ket\pm$ states behave as solitons.
They do not create or annihilate with each other or deform from one to the other. 
This elasticity suggests that the model is integrable\footnote{The integrability can be predicted also from the symmetry of the (bulk) Hamiltonian not being limited to the finite-dimensional $SU(2)\times U(1)_i$ algebra but extended to the infinite-dimensional Yangian algebra $Y(SL(2))$. }.
Indeed, the Hamiltonian can be constructed (up to the coupling $-1/2\lambda$ factor) from the R-matrix 
\begin{widetext}
\begin{align}\label{eq:r-straight}
    &R(\mu)
    = -i \begin{pmatrix}
      \sh (\mu +\eta) & 0 & 0 & 0 & 0 & 0 & 0 & 0 & 0 \\
 0 &\sh(\mu ) & 0 &\sh(\eta ) & 0 & 0 & 0 & 0 & 0 \\
 0 & 0 &\sh(\mu ) & 0 & 0 & 0 &\sh(\eta ) & 0 & 0 \\
 0 &\sh(\eta ) & 0 &\sh(\mu ) & 0 & 0 & 0 & 0 & 0 \\
 0 & 0 & 0 & 0 &\sh(\mu + \eta) & 0 & 0 & 0 & 0 \\
 0 & 0 & 0 & 0 & 0 & 0 & 0 &\sh(\mu + \eta) & 0 \\
 0 & 0 &\sh(\eta ) & 0 & 0 & 0 &\sh(\mu ) & 0 & 0 \\
 0 & 0 & 0 & 0 & 0 &\sh(\mu + \eta) & 0 & 0 & 0 \\
 0 & 0 & 0 & 0 & 0 & 0 & 0 & 0 &\sh(\mu + \eta) 
    \end{pmatrix}, 
\end{align}
\end{widetext}
where $\sh$ is a shorthand for $\sinh$, with the spectral parameter $\mu$ and the ``anisotropy" $\eta = i\pi/2$.
We focus on this specific value of $\eta$ for our interest. 
The overall factor $-i$ is taken so that the R-matrix with the zero spectral parameter coincides with the permutation matrix. 
This R-matrix satisfies the Yang-Baxter (YB) equation
\begin{align}\label{eq:YB-R}
    &R_{12}(\lambda - \mu) R_{13}(\lambda) R_{23}(\mu) 
    =  R_{23}(\mu)R_{13}(\lambda)R_{12}(\lambda - \mu), 
\end{align}
and hence the spin chain is integrable. 
The Hamiltonian is obtained by expanding the transfer matrix associated to $R$ around $\mu=0$ as the spectral parameter (see for example \cite{Retore:2021wwh}).  
We compute the dispersion relation and the effective central charge of this massless model using the algebraic Bethe ansatz in Appendix~\ref{sec:three-letter-Bethe-ansatz}.

One can also see the integrability proving that this model is equivalent to free fermions~\cite{Katsura2023}. 
The Hilbert space can be factorized as 
\begin{align}
    \mc H = \left(\bigotimes_{j = 1}^L \mc H_F\right) \otimes 
    \mc H_{u,d}. 
\end{align}
Here, $\mc H_F$ is a fermionic Fock space at each letter site where the $\ket 0$ state corresponds to the $r$ letter and $\ket 1$ corresponds to either $u$ or $d$. 
$\mc H_{u,d}$ are spanned by the states describing the order of $u$ and $d$ appearing in the words. 
Since the letters $u$ and $d$ cannot pass through each other, the Hamiltonian acts nontrivially only on the fermionic Fock spaces as free-fermion hoppings and treats what we termed the knottiness as
a label for the different superselection sectors. 
This model (with a deformation by a chemical potential term) has appeared before in \cite{Alcaraz:1991ps, Alcaraz:1992zc, Gomez:1992nd, Maassarani:1997kon}\footnote{The authors thank Hosho Katsura for pointing out these references. } where it was
also shown to be integrable. 
Our important result is that the three-state spin chain with constraints can be converted to a nearest-neighbor spin chain without constraints, which happens to be a known integrable model.
We may call this the three-state representation of the XX spin chain as it has the massless free fermion description with the same effective dispersion relation as shown in Appendix~\ref{sec:three-letter-Bethe-ansatz}. 
This is a good indication that this framework can reproduce the correct continuum behavior, given that the two-letter and three-letter strings would be no longer distinguishable from each other in the continuum. 

\subsection{Four-letter strings \label{sec:four-letter}}
We have showed the integrability structures of the first-order perturbation in the two-letter and three-letter sectors in the previous sections. 
Here, we list some phenomena in the most general sectors in $2+1$-D, the four-letter sectors, and show the evidence of non-integrability.  

\begin{figure}
    \centering
    \subfloat[]{\label{fig:loop-scattering1} 
    \includegraphics[width=0.3\linewidth]{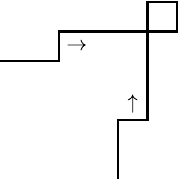}
    }
    \subfloat[]{\label{fig:loop-scattering2} 
    \includegraphics[width=0.3\linewidth]{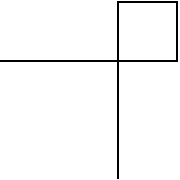}
    }
    \subfloat[]{\label{fig:loop-scattering3} 
    \includegraphics[width=0.3\linewidth]{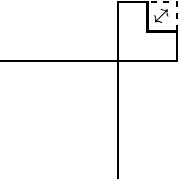}
    }
    \caption{A four-letter string configuration with a loop at a corner (non-trivial secondary knottiness). The defect scattering coming into the loop ((a)) may transform to a fluctuation inside of the loop ((b) and (c)). This is obviously a non-integrable scattering process, as the additional state only appears when two excitations from different sides of the loop collide simultaneously with the loop. }
    \label{fig:loop-scattering}
\end{figure}

We can find a non-integrable scattering process only adding a few fourth ($l$) letters in the three-letter configuration. 
Consider the configuration with a loop at a corner of a long string stretching horizontally on one side and vertically on the other side due to one $l$ configuration as Fig.~\ref{fig:loop-scattering}. 
The scatterings far from the loop are the same as the three-letter case, which means they are elastic, but once they enter the region close enough to the loop, they become nontrivial. 
First, it should be noticed that the loop itself is a rigid structure on top of the horizontal plus vertical line.
The constraints forbid any of the letters present to  change location as such a process would violate one of the knottiness constraints.
Such rigidity only appears because of the projectors and what we called secondary knottiness. Without the projectors, one can find a decay path of the loop so that the loop can decay into moving excitations.

Now consider two additional defect excitations approaching the rigid loop as Fig.~\ref{fig:loop-scattering1}. 
If they collide with the loop at the same time, the size of the loop becomes larger (Fig.~\ref{fig:loop-scattering2}), allowing fluctuation inside of the loop (Fig.~\ref{fig:loop-scattering3}). 
This kind of scattering processes is inelastic, suggesting the non-integrability of the four-letter sectors. The reason for this inelasticity is that the two defects must reach the rigid loop simultaneously in order to find the extra excitation of the loop. Each of the defects on its own would be reflected back with some phase shift. This counts effectively as a three body interaction that does not factorize.

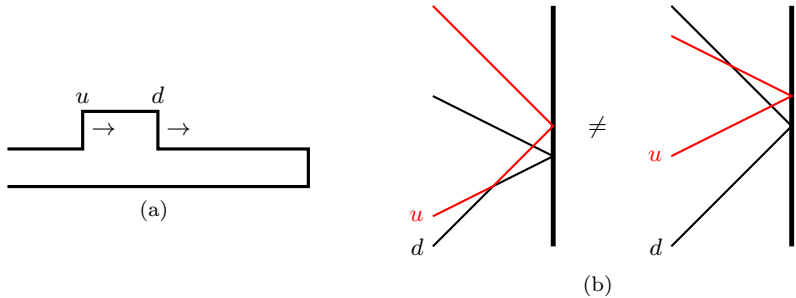
\begin{figure}
    \centering
    \subfloat[]{\label{fig:fold-scattering} \begin{tikzpicture}[baseline = (current bounding box.center)]
        \draw[very thick] (0, 0) --(1, 0) -- (1, 0.5) -- (2, 0.5) -- (2, 0) -- (4, 0) -- (4, -0.5) -- (0, -0.5); 
        \node[right] at (1, 0.25) {$\rightarrow$}; 
        \node[right] at (2, 0.25) {$\rightarrow$};
        \node[above] at (1, 0.5) {$u$}; 
        \node[above] at (2, 0.5) {$d$};
    \end{tikzpicture}}
    \hspace{1cm} 
    \subfloat[]{\label{fig:fold-scattering-boundary} 
     \begin{tikzpicture}[scale = 0.8, baseline = (current bounding box.center)]
        \draw[line width = 0.2em] (0, 0) -- (0, 4); 
        \draw[thick] (-2, 0) node[left] {$d$} -- (-1, 1) -- (0, 1.5) -- (-2, 2.5); 
        \draw[thick, red] (-2, 0.5) node[left, red] {$u$} -- (-1, 1) -- (0, 2) -- (-2, 4); 
    \end{tikzpicture}
    \quad $\neq$ \quad 
    \begin{tikzpicture}[scale = 0.8, baseline = (current bounding box.center)]
        \draw[line width = 0.2em] (0, 0) -- (0, 4); 
        \draw[thick] (-2, 0) node[left] {$d$} -- (0, 2) -- (-2, 4); 
        \draw[thick, red] (-2, 1.5) node[left, red] {$u$} -- (0, 2.5) -- (-2, 3.5); 
    \end{tikzpicture}
    }
    \caption{(a) The two kinks ($u$ and $d$) propagating on a folded string. $d$ can pass through the folding point (the $d$ edge between the $r$ letters on the top and the $l$ letters on the bottom), while $u$ cannot pass through it but has to bounce back. (b) The two different scattering processes on this folded string depending on whether the two kinks can interact on the top side (right) or not (left) before $d$ passes the folding. }
    \label{fig:fold-scattering-four-letters}
\end{figure}

We may consider another inelastic scattering process which can be translated to boundary conditions violating the boundary Yang-Baxter equation \footnote{This is a similar technique to the one used in \cite{Berenstein:2004ys} to show generic non-integrability for deformations of ${\cal N}=4 $ SYM. }.
Think of a folded string as Fig.~\ref{fig:fold-scattering}. The position of the fold is rigid because of the knottiness constraints. In this case, we can say that the position of the fold is fixed by where we require the padding to be located. In that sense, as a system with three letters it corresponds to a configuration with all fermion sites occupied in the free fermion picture. 
Since the fermion sector is the only sector involving dynamics in the first order, this state is rigid. 

Now add two kinks in opposite vertical directions ($u$ and $d$) propagating on it towards the right, and the kink that is behind (let us say $u$) has a larger initial momentum than the other kink ($d$). 
The scattering processes qualitatively differs depending whether the two kinks interact on the top side (consisting of the $r$ background letters) of the folding or not. 
If the $u$ kink catches up with $d$ on the top side of the folding, then the two kinks reflect carrying each other's momentum. 
On the other hand, if the $d$ kink reaches the fold first, it can get past the fold easily as our map to the three-letter padded XX model would indicate. If that happens, the kinks do not interact with each other at all since the $u$ kink cannot pass the folding: it must be reflected back always. 

Regarding the folding point as the boundary and gluing the top and bottom sides together (just as the folding trick for an interface being mapped to a boundary~\cite{Wong:1994np}), the two scattering processes can be depicted as Fig.~\ref{fig:fold-scattering-four-letters}. 
In this picture, both processes where a $d$ kink passes through the $d$ folding point and a $u$ kink bouncing back at the folding point are elastic reflections at the boundary.
In such a reflection diagram they can interact in two possible ways, either a reflection (when they are on the same side of the folding) or an identity (they are on opposite sides). 
This obviously violates the boundary Yang-Baxter equation given that the final states are physically distinct. This is therefore a violation of integrability 
in the Yang-Baxter sense.  

Another potential fact hinting at the non-integrability is that the boost operator, which is naively defined as 
\begin{align}
    B = \Pi \Tilde B \Pi = \sum_\ell \ell \pi_\ell \Tilde h_\ell \pi_\ell,
\end{align}
does not coincide with the boost operator of the integrable Hamiltonian for the three-letter configurations Eq.~\eqref{eq:H-straight}: it does not generate the infinite set of conserved charges that one expected from the integrability of the three-letter problem. 
The basis change that pads the words correctly Eq.~\eqref{eq:3-letter-padding} does not work for the general four-letter sectors, causing it to violate the possible integrability of the spin chain. Again, this is a violation in that the system does not seem to be solvable by Bethe Ansatz methods. Remember also that all of this structure arises exactly because of the projectors, as without the projectors the nearest neighbor spin chain is integrable.

\section{Rotational symmetry restoration\label{sec:rot-sym-restoration}}
One way to validate our scheme is to see if it reproduces the phenomena expected in the continuum limit. 
In the continuum, an infinitely long flux tube as we consider here is expected to be delocalized~\cite{Luscher:1980ac} (and references therein) due to by the quantum fluctuation of the Goldstone bosons caused by the breaking of the Poincar\'e symmetry. 
This does not hold for the lattice theory since the Poincar\'e symmetry is already broken to be discrete, and there indeed exist a stable flux tube sector in the strong coupling limit. 
As we weaken the coupling, there exists a so-called roughening transition point where the long flux tube delocalizes. 
Below this point, we expect our effective string description well reflects the behavior of the flux tube in continuum, including the restoration of the continuous Poincar\'e symmetry in the bulk YM theory. 

We confirm this symmetry restoration by estimating the roughening transition in the two-letter and three-letter sectors using the degenerate corrections argued in~Sec.~\ref{sec:two-letter-1st} and \ref{sec:three-letter-1st}. 
Fortunately, we found that both Hamiltonians are integrable, and hence the exact calculations are feasible. 

\begin{figure}[t]
    \centering
    \begin{tikzpicture}[scale = 2, baseline = (current bounding box.center)]
        \draw[step=0.5,gray,very thin] (-1.2,-0.7) grid (1.2,1.2);
        \draw[very thick] (-1.2, 0) -- (0, 0)--(0, 0.5)--(1.2, 0.5);
    \end{tikzpicture}
    \caption{A single kink on a straight string. }
    \label{fig:kink}
\end{figure}
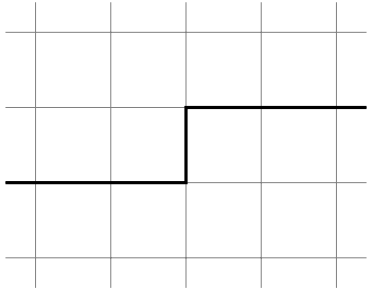

One of the possible approaches is to identify the roughening transition point as the point where a single kink looking like Fig.~\ref{fig:kink} becomes massless. 
In the strong coupling limit, such a string state with a kink costs extra energy exactly by the amount of the string tension from the energy of the fully straight string given that the energy is proportional to the length of the string. 
As the coupling becomes weaker and the single kink becomes massless, there is no longer a cost for the string to go to the vertical direction.
This leads to roughening of the string. 
This transition can be understood from the spin chain point of view by considering the unperturbed and perturbation parts of the Hamiltonian all together. 
The state under consideration consists of a number of $r$'s except there is one $u$. 
In the two-letter sector with the XX spin chain description (Sec.~\ref{sec:two-letter-1st}), this corresponds to the correction of the energy of a state with only one upspin, or a $-(L-2)$ magnetization state. 
In the basis of the first-order Hamiltonian in the three-letter sector (Sec.~\ref{sec:three-letter-1st}), the unperturbed part can be added to the first-order Hamiltonian as a chemical potential term $ \frac{\lambda}{2} \sum_{j=1}^L i_j^3$, i.e. the energy cost of the vertical kinks. 
In the strong-coupling limit, this unperturbed part dominates and hence the theory is massive, whereas the perturbation term Eq.~\eqref{eq:H-straight} dominates in the weak-coupling limit, which is massless. 
So, we expect the phase diagram of this model to have two regions, the massive phase and the massless phase, which are connected by a Berezinskii–Kosterlitz–Thouless (BKT)-like phase transition. 
This BKT-like transition point is identified as the roughening transition point. 
The transition in this class has been observed by studying the Wilson/Polyakov line for example in the discrete $\mZ_2$ lattice gauge theory in~\cite{Juge:2001mj, Juge:2001rb}. Other  setups for the electric flux roughening transition in $2+1$ D have been done in \cite{DiMarcantonio:2025cmf}.

Actually, it is possible to directly observe the restoration of the spatial rotational symmetry for the two-letter string case as discussed in~\cite{Kogut1981}. 
The angular dependence of the potential, which can be translated as the ground state energy in each magnetization sector of the XX model, should disappear once the symmetry is restored. 

We estimate the values of the coupling $\lambda$ restoring the rotational symmetry for these two cases by extrapolating from the strong-coupling expansions.

\subsection{Roughening point from the two-letter strings}
\subsubsection{First-order correction to the kink mass}
As we discussed earlier, the massive/massless transition of the kink mass corresponds to the roughening transition of the string if it occurs. 
In the strong coupling limit, the mass of the kink is exactly the energy of the single link excitation or the string tension, which is $\sigma = \lambda/2$. 
So, in the strong coupling region, the kink mass is nonzero, and this nonzero mass prohibits the string from extending in the vertical direction. 
On the other hand, if such a transition happens, and the kink mass becomes zero, the extension to the vertical direction no longer costs energy; this is what is called the roughening of the string. 
We calculate the value of the kink mass with the first-order correction to extrapolate the value of $\lambda$ giving the massless kink. 
A state with a single kink can be interpreted as a state with one upspin in the XX spin chain description. 
It is diagonalized as a free fermion theory 
as 
\begin{align}
    H_{\mrm{XX}} = -\frac{1}{\lambda} \sum_{k = -L/2+1}^{L/2} \Lambda_k \hat a_k^\dagger \hat a_k, \quad \Lambda_k = \cos(\frac{\pi k}{L})
\end{align}
in the thermodynamic limit (details in Appendix~\ref{sec:diagonal-dispersion}), and the operator counting the number of upspins is exactly the fermion number operator. 
Hence, the state of our interest has exactly one fermionic excitation with energy of $ -\frac{1}{\lambda} \Lambda_k$. 
What we need to consider is the state consisting of a stationary kink, which is given by setting $k = 0$ corresponding to zero momentum. 
The corrected kink mass is 
\begin{align}\label{eq:roughening-two-letter}
    M_{\text{kink}} = \frac{\lambda}{2} - \frac{1}{\lambda} + \mc O\left(\frac{1}{\lambda^2}\right). 
\end{align}
This becomes zero when $\lambda = \sqrt{2}$ which we identify as the roughening transition point up to the first order in the strong coupling expansion.

\subsubsection{Direct first-order estimation of the symmetry restoration}
\begin{figure}
    \centering
    \includegraphics[width=\linewidth]{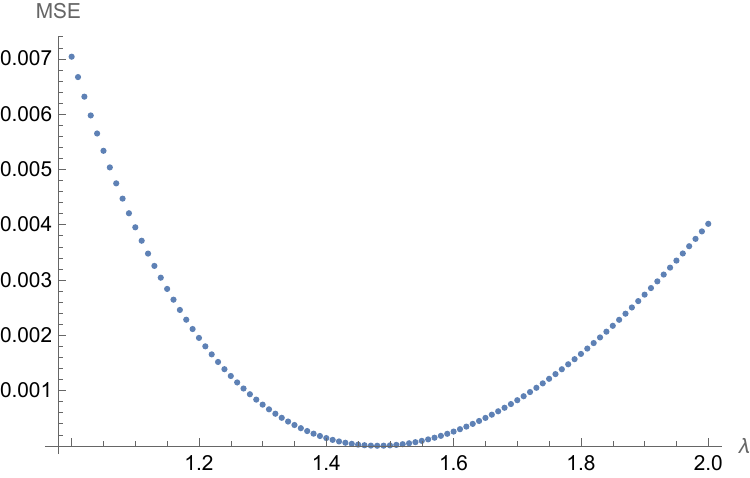}
    \caption{The mean squared errors of $E$ for different values of $\lambda$. It is minimized to be almost zero around $\lambda \approx 1.5$. }
    \label{fig:mse-diagonal}
\end{figure}

As mentioned above, we can directly observe the symmetry restoration in the two-letter string sector, following~\cite{Kogut1981}.
The angular dependence is the choice of the combination $(n, L-n)$ for the numbers of the two letters for the first-order corrections, which is exactly equivalent to choosing a certain magnetization sector of the XX model as we saw in Sec.~\ref{sec:two-letter-1st}. 
Let us say the lowest $n$ fermi levels are filled, which means the state is in the corresponding magnetization $M = n$ sector in the spin description, or having $n$ vertical and $L - n$ horizontal edges. 
The string state in the lowest energy in the $n$ sector has the energy 
\begin{align}
    E 
    & = \frac{\lambda}{2} L -\frac{1}{\lambda} \sum_{k = 0}^{n-1} \cos(\frac{\pi k}{L})
    = \frac{\lambda}{2} L -\frac{1}{2\lambda}\left(1 + \csc(\frac{\pi}{2L})\sin(\frac{\pi (2n-1)}{2L})\right) .  
\end{align}
In the continuum, we can write the parameters in the polar coordinates; defining $\rho$ as the physical distance of the endpoints of the string and $\theta$ as their angle from the horizontal axis, the number of the horizontal edges is $\ell = \rho \cos\theta$, and the string length is $L = \rho \cos\theta + \rho \sin\theta$. 
Now, the ground state energy is expressed as 
\begin{align}
    E 
    &= 
    \rho (\cos\theta + \sin\theta)\left(\frac{\lambda}{2} - \frac{1}{\pi\lambda}\sin(\frac{\pi }{1 + \tan\theta})\right) 
    + \mc O(\rho^0). 
\end{align}
Its derivative by $\theta$ at $\theta = 0$, corresponding to a string consisting almost all $r$'s, becomes $0$ at $\lambda = \sqrt{2}$ in the leading order of $\rho$, which is exactly what we estimated by considering the first-order correction to the kink mass in the last section. 
Indeed, $E $ is almost constant around $\lambda \approx 1.5$, as we can see by numerically calculating where $E$ becomes the closest to a constant as a function of $\theta$ (Fig.~\ref{fig:mse-diagonal}). 

\subsection{Roughening point from the three-letter strings}
Let us estimate the roughening transition point in the three-letter sector as well. 
Since the two-letter strings and the three-letter strings would be no longer distinguishable from each other once we reach continuum, the transition point should coincide if expansions in all orders were available.
Indeed, we will demonstrate that the extrapolated value of $\lambda$ in the first-order calculation already matches with the result from the two-letter sector calculation. 
We also carry out the second-order calculation to confirm two things.
First is to estimate the point when the string tension becomes negative, which we estimate to happen after the roughening phase transition.  
Second is that the extrapolated transition point does not deviate too far from the first-order result and hence our perturbative expansions are providing meaningful estimates. 

\subsubsection{First-order corrections}
Let us compute the first-order correction to the kink mass, using the Hamiltonian Eq.~\eqref{eq:H-straight}. 
The single kink state lives in the sector of $U(1)_i$ charge of $i = 1$. 
This corresponds to acting a single ($n=1$) $B$ matrix on the Bethe reference state $\ket \Omega$ (details in Appendix~\ref{sec:three-letter-Bethe-ansatz}).
The energy and momentum spectra we found in Eq.~\eqref{eq:energy-spectrum} lead to the dispersion relation for $n=1$ to be 
\begin{align}
    E_{1, \ell}^{(1)} &= -\frac{1}{\lambda} \cos(p_\ell^{(1)})
\end{align}
with $p^{(1)}_\ell = 2\pi \ell/L$ and $\ell = 0, 1, ..., L-1$. 
Here, we recovered the coupling factor $-1/2\lambda$. 
This is the first-order correction to the single kink mass which is overall $
    M_{\text{kink}} = \frac{\lambda}{2} -\frac{1}{\lambda}\cos(\frac{2\pi \ell}{L}) + \mc O(1/\lambda^2)$. 
Since we want to consider the stationary kink with $p_\ell^{(1)} = 0$, we have 
\begin{align}
    M_{\text{kink}} = \frac{\lambda}{2} - \frac{1}{\lambda} + \mc O\left(\frac{1}{\lambda^2}\right)
\end{align}
This is exactly the expression we obtained in the two-letter sector (Eq.~\eqref{eq:roughening-two-letter}). 
Again, the roughening point is estimated as $\lambda = \sqrt{2}$.

\subsubsection{Second-order corrections}

We now want to consider second order corrections. Let us start with the corrections to the string tension for the horizontal state (in letters $rrrrrrr\dots$).
This state is rigid and can be considered a ground state with energy $E_0=\sigma L $. Notice that there is no other state with the same quantum numbers that is degenerate with it and that the expectation value of the perturbation plaquette vanishes on this state.
When we add perturbations to second order perturbation theory, we need to compute
\begin{equation}
  \Delta E= \sum_j \frac{V_{ij}^2}{E_0-E_j}
\end{equation}
where $V_{ij}$ are the off-diagonal matrix elements. Since $E_0<E_j$, the correction will be negative.

There are two types of moves allowed by inserting  a plaquette.
These are of type C, and another one of type B that joins a closed rigid loop at any location by intertwining the lines. We will call this operation $B'$.

The type C moves take a letter $r\to urd$ or $r\to dru$ and adds a kink-antikink pair. The matrix element is $1$, times $(N/2\lambda)\frac 1N$ for each of them. The difference in energy is $2\sigma$ as the length changes by $2$. 
The contribution to the energy is therefore
\begin{equation}
   \Delta E_{\text{C}}= \sum_{\ell = 1}^L \frac{2}{-2 \sigma}\frac 1 {2^2\lambda^2} =  -\frac{1}{2 \lambda^3}L, 
\end{equation}
and we notice that this is proportional to the length, so the correction to the tension is
\begin{equation}
    \Delta \sigma_{\text{C}}= -\frac{1}{2 \lambda^3}. 
\end{equation}

The second contribution requires adding (for example) a loop $uldr$ at some position, which is a loop on top of the string. We can also add the loop $ruld$, one step to the left. Both of these cause the same transition and therefore the combinatorial factor associated to this term is actually a factor of $2$. This is exactly the combinatorial factor that appeared in Eq.~\eqref{eq:overlap-twist-loop}. The same $1/{2\lambda}$ is attached to plaquette, and the denominator is $-4\sigma$ rather than $-2\sigma$.
We get this way that (when we sum over up and down)
\begin{equation}
    \Delta \sigma_{{\text{B}}'}= \frac{-2}{4 \sigma}\left[\frac{2}{2 \lambda}\right]^2= -\frac 1{\lambda^3}
\end{equation}
Summing the two terms, we get that
\begin{equation}
    \Delta \sigma_{\text{C}+\text{B}'} = -\frac{3}{2\lambda^3}
\end{equation}
Notice that $\sigma$ wants to become negative (tensionless) when $\lambda= 3^{1/4}\sim 1.31$, which is below the value where the mass of the kink becomes zero for the first time. That means that at least at this order, the massless kink that corresponds to the roughening transition comes first.

Let us further consider the second-order corrections to the single kink energy. 
The on-diagonal contribution to the single kink state $\ket x$ is 
\begin{align}
    E_\ell^{(2)} = \sum_{k \not\in D_\text{1-kink}} \frac{|\mel{k^{(0)}}{V}{x}|^2}{E^{(0)}_{D_{\text{1-kink}}}- E^{(0)}_k}, 
\end{align}
where $D_{\text{1-kink}}$ is the degenerate sector spanned by all of the one-kink states. 
$x$ denotes the position of the kink on the straight string. 
The nonzero contributions come from the states with a deformation made by a single plaquette from a one-kink state. 
In order to combine with the first-order correction, to be more precise, we actually have to take the state under consideration to be the eigenstate of the first-order correction matrix corresponding to a stationary kink, which would be a linear combination of $\ket x$. 
However, since the second-order corrections to single kink states are all equal with each other (at least in the region away from the string endpoints), we ignore this subtlety and simply consider a kink at a fixed position.

Most of the contributions are the same as the second-order corrections to the straight string without any kink, except for the corrections due to plaquette actions around the kink, which we will consider below. There are three things we need to take care of.

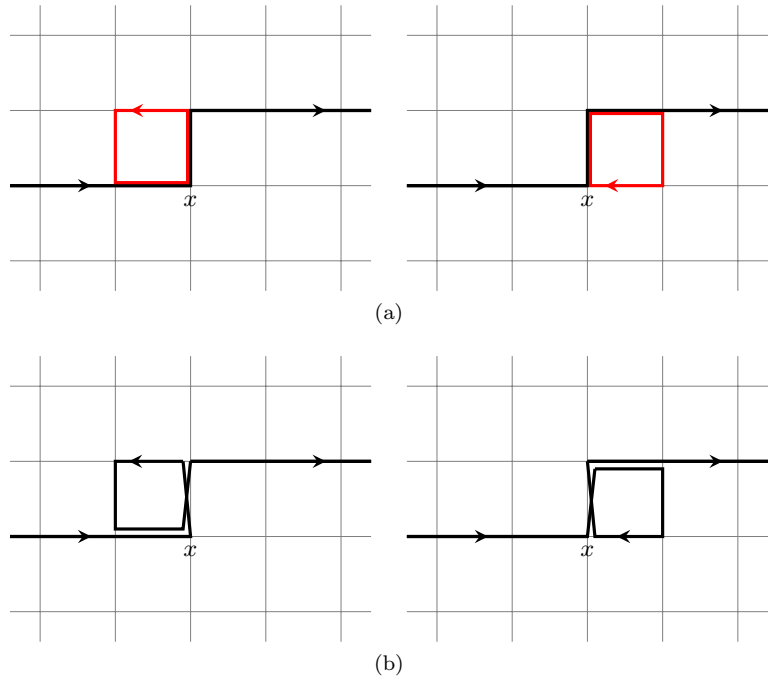
\begin{figure}[t]
    \centering
    \subfloat[]{\label{fig:kink-2nd-b}
    \begin{tikzpicture}[scale=2, baseline = (current bounding box.center)]
        \draw[step=0.5,gray,very thin] (-1.2,-0.7) grid (1.2,1.2);
        \draw[very thick] (-1.2, 0) edge [mid arrowpos={black}{0.45}] (0, 0)  -- (0, 0) node [below] {\small $x$} -- (0, 0.5) edge [mid arrowpos={black}{0.75}] (1.2, 0.5) --(1.2, 0.5);
        \draw[very thick, red, mid arrowpos ={red}{0.2}] (-0.02, 0.5) -- (-0.5, 0.5) -- (-0.5, 0.02) -- (-0.02, 0.02) -- cycle;
    \end{tikzpicture}
    \quad 
    \begin{tikzpicture}[scale=2, baseline = (current bounding box.center)]
        \draw[step=0.5,gray,very thin] (-1.2,-0.7) grid (1.2,1.2);
        \draw[very thick] (-1.2, 0) edge [mid arrowpos={black}{0.45}] (0, 0)  -- (0, 0) node [below] {\small $x$} -- (0, 0.5) edge [mid arrowpos={black}{0.75}] (1.2, 0.5) --(1.2, 0.5);
        \draw[very thick, red, mid arrowpos ={red}{0.7}] (0.02, 0.48) -- (0.5, 0.48) -- (0.5, 0) -- (0.02, 0) -- cycle;
    \end{tikzpicture}
    }\\
    \subfloat[]{\label{fig:kink-overlap-b}
    \begin{tikzpicture}[scale=2, baseline = (current bounding box.center)]
        \draw[step=0.5,gray,very thin] (-1.2,-0.7) grid (1.2,1.2);
        \draw[very thick] (-1.2, 0) edge [mid arrowpos={black}{0.45}] (0, 0)  -- (0, 0) node [below] {\small $x$} -- (-0.05, 0.5);
        \draw[very thick] (0, 0.5) edge [mid arrowpos={black}{0.75}] (1.2, 0.5) --(1.2, 0.5);
        \draw[very thick, mid arrowpos ={black}{0.2}] (-0.05, 0.5) -- (-0.5, 0.5) -- (-0.5, 0.05) -- (-0.05, 0.05) -- (0, 0.5);
    \end{tikzpicture}
    \quad 
    \begin{tikzpicture}[scale=2, baseline = (current bounding box.center)]
        \draw[step=0.5,gray,very thin] (-1.2,-0.7) grid (1.2,1.2);
        \draw[very thick] (-1.2, 0) edge [mid arrowpos={black}{0.45}] (0, 0)  -- (0, 0) node [below] {\small $x$} -- (0.05, 0.45);
        \draw[very thick] (0, 0.5) edge [mid arrowpos={black}{0.75}] (1.2, 0.5) --(1.2, 0.5);
        \draw[very thick, mid arrowpos ={black}{0.65}] (0.05, 0.45) -- (0.5, 0.45) -- (0.5, 0) -- (0.05, 0) -- (0, 0.5);
    \end{tikzpicture}
    }
    \caption{(a) The type~B deformations to a string state with a single kink at position $x$ which have a nonzero overlap with another states depicted in (b), in the leading-order. The contributions from these overlaps do not appear in the corrections to the no-kink state. }
    \label{fig:2nd-kink-typeb}
\end{figure}

First, one needs to consider the type~B$'$ deformations on the kink (Fig.~\ref{fig:2nd-kink-typeb}). 
If we look at a word that goes $rur$ and need to insert a loop, there are two choices for replacing $u
\to u(rdlu) $, with a loop that twists to the right, which would add a word to the right of $u$, or we could add the word $(urdl)u$ to the left. Moreover, this interferes with adding $ur(dlur)$ with the next letter. The combinatorial factor for the off-diagonal element is therefore $3$, compared to $2$ when we have just a string of letters of type $r$. We get a similar contribution from a loop that loops to the left (eg, $u\to uldru$).
We get
\begin{equation}
\Delta E_{\text{B}'}= 2 \times \frac {9}{-4 \sigma}\frac{1}{(2\lambda)^2}
\end{equation}
However, we are double counting the contribution to the tension of the string from type $B'$ moves (if we consider them independent) acting on the $r$ edge overlapping with the loop, so we need to subtract the correction to the string tension correctly from the kink contribution. We arrive at
\begin{equation}
\Delta E_{\text{B}'}(kink)= 2 \times \frac {9}{-4 \sigma}\frac{1}{(2\lambda)^2}
- 2 \times \frac {4}{-4 \sigma}\frac{1}{(2\lambda)^2}
=
-\frac{5}{8\lambda^3}. 
\end{equation}

\begin{figure}
    \centering
    $\vcenter{\hbox{\includegraphics[width=0.3\linewidth]{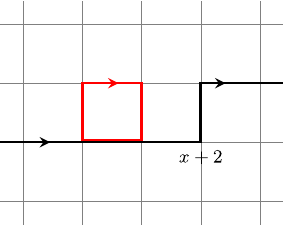}}}
 \,\rightarrow \, 
\vcenter{\hbox{\includegraphics[width=0.3\linewidth]{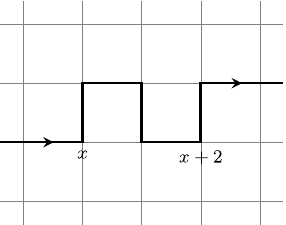}}}
\, \leftarrow \, 
\vcenter{\hbox{\includegraphics[width=0.3\linewidth]{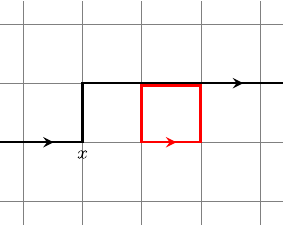}}}
    $
    \caption{Type~C deformations can act on different kink states (left ($urr$) and right ($rru$) and result in the same intermediate state ($urdru$). This lead to off-diagonal contribution in the second order. }
    \label{fig:off-diagonal-typeC}
\end{figure}

\begin{figure}[t]
    \centering
    \subfloat[]{\label{fig:kink-2nd-c}
    \begin{tikzpicture}[scale = 2, baseline = (current bounding box.center)]
        \draw[step=0.5,gray,very thin] (-1.2,-0.7) grid (1.2,1.2);
        \draw[very thick] (-1.2, 0) edge [mid arrowpos={black}{0.45}] (0, 0)  -- (0, 0) node [below] {\small $x$} -- (0, 0.5) edge [mid arrowpos={black}{0.75}] (1.2, 0.5) --(1.2, 0.5);
        \draw[very thick, red, mid arrowpos ={red}{0.85}] (-0.5, 0) -- (-0.5, 0.5) -- (0, 0.5);
        \draw[very thick, red, dashed] (-0.5, 0) -- (0, 0) -- (0, 0.5);
    \end{tikzpicture}
    \quad 
    \begin{tikzpicture}[scale = 2, baseline = (current bounding box.center)]
        \draw[step=0.5,gray,very thin] (-1.2,-0.7) grid (1.2,1.2);
        \draw[very thick] (-1.2, 0) edge [mid arrowpos={black}{0.45}] (0, 0)  -- (0, 0) node [below] {\small $x$} -- (0, 0.5) edge [mid arrowpos={black}{0.75}] (1.2, 0.5) --(1.2, 0.5);
        \draw[very thick, red, mid arrowpos ={red}{0.4}] (0, 0) -- (0.5, 0) -- (0.5, 0.5);
        \draw[very thick, red, dashed] (0, 0) -- (0, 0.5) -- (0.5, 0.5);
    \end{tikzpicture}
    }\\
    \subfloat[]{\label{fig:no-kink-2nd-c}
    \begin{tikzpicture}[scale = 2, baseline = (current bounding box.center)]
        \draw[step=0.5,gray,very thin] (-1.2,-0.7) grid (1.2,1.2);
        \draw[very thick] (-1.2, 0) edge [mid arrowpos={black}{0.45}] (0, 0) -- (0, 0) node [below] {\small $x$} (0,0) edge [mid arrowpos={black}{0.75}] (1.2, 0) --(1.2, 0);
        \draw[very thick, red, , mid arrowpos={red}{0.55}] (-0.5, 0) -- (-0.5, 0.5) -- (0, 0.5) -- (0,0);
        \draw[very thick, red, dashed] (-0.5, 0) -- (0, 0);
    \end{tikzpicture}
    \quad 
    \begin{tikzpicture}[scale = 2, baseline = (current bounding box.center)]
        \draw[step=0.5,gray,very thin] (-1.2,-1.2) grid (1.2,0.7);
        \draw[very thick] (-1.2, 0) edge [mid arrowpos={black}{0.45}] (0, 0) -- (0, 0) node [above] {\small $x$} (0,0) edge [mid arrowpos={black}{0.75}] (1.2, 0) --(1.2, 0);
        \draw[very thick, red, mid arrowpos={red}{0.6}] (0, 0) -- (0, -0.5) -- (0.5, -0.5) -- (0.5,0);
        \draw[very thick, red, dashed] (0, 0) -- (0.5, 0);
    \end{tikzpicture}
    }
    \caption{(a) The type~C deformations to a string state with a single kink at position $x$ which do not contribute to the second-order corrections since they are already taken into account in the first-order corrections. 
    (b) The corresponding type~C deformations to the no-kink state which do contribute to the second-order corrections. \label{fig:second-order-typeC}}
\end{figure}
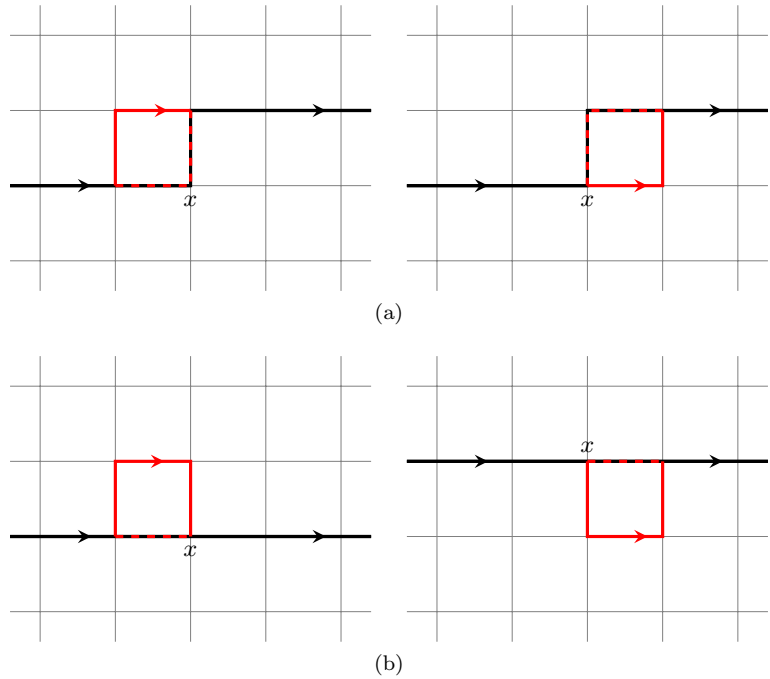

We also need to remove the C moves that are absent from the string tension but already taken care of in the first-order corrections for the kink (Fig.~\ref{fig:second-order-typeC}).
Moreover, there is one more type~C contribution  that is not obvious. This also arise from C type moves, but it is off-diagonal. The idea is that the configurations of type $urdru$ can be accessed in two ways. Starting from $rru\rightarrow (urd)ru$ and $urr\rightarrow ur(dru)$ where we have put in parenthesis the replacement of the letter $r$ by the $C$ type move (one upwards and the other one downwards). This has energy denominator $-2\sigma$ and produces an effective off-diagonal substitution $rru\leftrightarrow urr$ which corresponds to a next to nearest neighbor hop (Fig.~\ref{fig:off-diagonal-typeC}, see also \cite{Kogut1981}).
Since the on-diagonal corrections to these two states are identical, this off-diagonal contribution simply produces an extra term to the energy eigenvalue as 
\begin{equation}
\Delta E_{\text{O.D.}}= \sum_j \frac{V_{ij}V_{jk}}{-2 \sigma}= 
\frac{1}{-2\sigma}\frac{1}{(2\lambda)^2} (2\cos(2p_\ell)).
\end{equation}
This  modifies the momentum dependence  of the dispersion relation. 
We need to evaluate it at $p_{\ell}=0$, so.
\begin{equation}
\Delta E_{\text{O.D.}}= -\frac{1}{2\lambda^3}
\end{equation}

Lastly, we need to include the nontrivial corrections due to overlaps with the type~A$'$ deformations at either one of the two corners of the kink. 
This deformation corresponds to inserting either a single clockwise or counter-clockwise loop before or after the kink $u$. For example the two moves $u\to uuldr$, or $u\to ulurd$ generate corner terms with additional loops wit two possible orientations.
The matrix elements of the perturbation $V_{ij}$ for these overlaps are again $(N/2\lambda)\frac{1}{N}$, where the extra $1/N$ factor comes from the topology change at the shared corner. 
The energy denominator is $-4\sigma$ due to the additional single plaquette loop, and the correction is 
\begin{align}
    \Delta E_{\text{A}'}(kink) = 2\times 2 \times \frac{1}{-4\sigma}\frac{1}{(2\lambda)^2} = -\frac{1}{2\lambda^3}. 
\end{align}

These contributions give the energy correction in total as 
\begin{align}
&\Delta E(kink) \nn  &= 
\Delta E_{\text{B}'}(kink) - \Delta \sigma_{\text{C}} + \Delta E_{\text{O.D.}} +  \Delta E_{\text{A}'}(kink)\nn
&= 
-\frac{5}{8\lambda^3}
 +\frac 1{2\lambda^3}-\frac{1}{2\lambda^3} -\frac{1}{2\lambda^3}\nn 
&=  -\frac{9}{8\lambda^3}. 
\end{align}
Hence, the kink mass with the corrections up to the second order dispersion is 
\begin{align}
    M_{\text{kink}} = \frac{\lambda}{2} - \frac{1}{\lambda} -\frac{7}{8\lambda^3} + \mc O\left(\frac{1}{L} + \frac{1}{\lambda^4}\right). 
\end{align}
This becomes zero when $\lambda = (1 + \sqrt{13}/2)^{1/2}\approx 1.67$.

\section{Higher dimensions\label{sec:higher-dimensions}}
As noted in section~\ref{sec:setup}, our word representation of confining strings is valid regardless of the spacetime dimensions. 
This indicates that it is straight-forward to extend it to higher dimensions; it adds additional letters to the set of letters we already have and with the additional zigzag constraints and the  notion of the knottiness (that for each pair of back and forth letters the order cannot change). 
Let us comment on some subsectors in $3+1$-dimensions.

\subsection{Three-letter sector in $3+1-D$}
Consider $3+1$-dimensions, and string configurations with three letters that describe the edge excitations heading towards positive $x, y, z$ directions. 
It is an analogue of the two-letter sector in $2+1$-D, given that the sector of these configurations do not require any projection, and the spin chain Hamiltonian describing the first-order corrections is an operator in $(\mC^3)^{\otimes L}$: 
\begin{align}
    H &= -\frac{1}{2\lambda}  \sum_{j = 1}^L (e^1_j f^1_{j+1} + f^1_j e^1_{j+1} + e^2_j f^2_{j+1} + f^2_j e^2_{j+1} + e^3_j f^3_{j+1} + f^3_j e^3_{j+1} )
\end{align}
where the operators $e^a_i$ and $f^a_i$ are the $\mathfrak{sl}(3)$ generators~\eqref{eq:gen-sl3}. 
This is exactly the bilinear-biquadratic spin-1 chain with 
\begin{align}
    H \propto \sum_{j = 1}^L \left(\cos\theta({\Vec S}_j\cdot {\Vec S}_{j+1}) + \sin\theta({\Vec S}_j\cdot {\Vec S}_{j+1})^2\right)
\end{align}
at $\theta = \pi/4$, i.e. the Uimin-Lai-Sutherland point, up to the chemical potential term. 
This point is integrable (see e.g.~\cite{Ambjorn:1999ei} for the exact form of the R-matrix), and more importantly, it can be effectively described as the $SU(3)_1$ Wess-Zumino-Witten model~\cite{Itoi1997}, which has a central charge of $c = 2$. 

\subsection{Four and more letter sectors in $3+1$-D}
\begin{figure}
    \centering
    \subfloat[]{\label{fig:xyx-3-body}
    \includegraphics[]{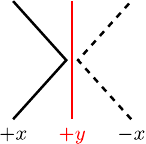}
    }\qquad 
    \subfloat[]{\label{fig:xyx-2-body-decomp}
     \includegraphics[]{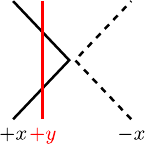}
    }
    \caption{An example of the three-body scattering on four-letter strings in $3+1$-D. The scattering process in~(a) where the $\pm x$ kinks and the $+y$ kink collide at the same time could be decomposed to two-body scatterings as in~(b) if there were no constraints. The zigzag constraints prohibit the direct interaction between $\pm x$ kinks so the reflection between them (the second two-body interaction in the decomposition~(b)) is not in the S-matrix data of the model, and hence the decomposition is not allowed. }
    \label{fig:four-letter-3-body}
\end{figure}

The next simplest case is the four-letter sector, which consists of string configurations with edges heading the positive or negative $x$-directions or the positive $y$- or $z$-directions. 
Although this is analogous to the three-letter sector in $2+1$-D, requiring only one type of projection that prohibits neighboring $\pm x$ edges, the situation is drastically more complicated. 
The basis change like Eq.~\ref{eq:3-letter-padding} would not work for this case, since now the $\pm x$ edges can have either an $+y$ edge or an $+z$ edge in between. 
The system also admits dynamics involving $+y$ and $+z$ edges, so we cannot simply discard one between the $\pm x$ edges. 
Actually, we should not even expect integrability for this sector, as one can see by regarding the $\pm x$ and $+y$ kinks as pseudo-particles on the vacuum of $+z$ edges or a long string in the $+z$-direction. 
Although the S-matrix for two-body scatterings between the $+x$ or $-x$ kink and the $+y$ kink is merely a permutation, three-body scatterings involving all three kinks cannot be decomposed into two-body scatterings as explained diagrammatically in Fig.~\ref{fig:four-letter-3-body}.
The zigzag constraint prohibits a three-body scattering like Fig.~\ref{fig:xyx-3-body} to be decomposed to the product of two-body scatterings as in Fig.~\ref{fig:xyx-2-body-decomp} since the $\pm x$ kinks are not allowed to be next to each other (their scatterings in the three-letter sector in $2+1$-D were special since they were allowed to be next to each other after the basis change Eq.~\eqref{eq:3-letter-padding}). 

The nonintegrability of sectors with more letters is clearer, since the strings contain at least two sets of edges heading parallel but in opposite directions similarly as the four-letter sector in $2+1$-D. 
The same argument as section~\ref{sec:four-letter} applies to these cases.

\section{Another Language\label{sec:another-language}}
So far we have investigated various subsectors of confining strings and their strong-coupling corrections. 
Even though there exist some subsectors where the projection due to the zigzag constraints is unnecessary,
it might be more convenient to describe the string configurations with a different convention to avoid the description with projections altogether.
If we think of the projections as a gauge constraint, this would be equivalent to finding the solutions to the constraint first, so that the dynamics becomes unconstrained in such a basis.

This is possible by expressing the words by the relative directions of the edges from each vertex relative to the incoming string segment. For example, for the strings in $2+1$-dimensions, the possible configuration for each edge is either straight ($\textsf s$), right ($\textsf r$), or left ($\textsf l$) relative to the incoming edge.  
Then, the zigzag symmetry constraint is implemented without any need of projections since the configurations going back and forth on the same edge are manifestly not included. 
In this language, one can also write the first-order corrections as the eigenvalues of a spin chain Hamiltonian, where each spin belongs to a  spin-1 chain, corresponding to the three possible relative directions. 
Instead of the projection, now the Hamiltonian includes next-to-nearest-neighbor interactions, since the letter after the two letters corresponding to the edges exchanged by the Type~C deformation is also affected. 
The Hamiltonian density with three-site interaction exchange letters as one of the following four pairs: 
\begin{align}
    \textsf{srs} \leftrightarrow \textsf{rlr}, \quad\textsf{srl} \leftrightarrow \textsf{rls}, \quad\textsf{sls} \leftrightarrow \textsf{lrl} , \quad \textsf{slr} \leftrightarrow \textsf{lrs}. \label{eq:exchange_dir}
\end{align}
This language can be translated back to the former $u,d,r,l$ language by assigning $\mZ_4$ values $0, 1, 3$ respectively to $\textsf{s, r, l}$ and summing these numbers from one end. 

The interesting fact is that in this form, the equation \eqref{eq:exchange_dir} also looks like a relatively simple set of rules.
The absence of projectors makes it in principle easier to implement.
This is favorable from the viewpoint of computational resources for (quantum) simulation. 
The former language naively requires a $4^4 = 256$ dimensional operator for each plaquette action taking the projection $\pi_\ell$ into account, while this language requires an operator with only $3^3 = 27$ dimensions. 
The $4^4$ requires $8$ qubits to implement, where the $27$ can be easily fit with $6$ qubits (each set of 3 letters can be built from $2$ qubits and leaving out one unwanted state). 

\begin{figure}
    \centering
    \def\a{0.7} 
\def\rt{1.7320508075688772} 
\def\Cols{3} 
\def\Rows{2} 

\subfloat[]{\label{fig:hexagon-original}
\includegraphics[width=0.45\linewidth]{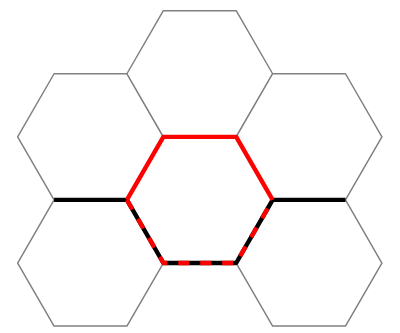}

} 
\,  
\subfloat[]{\label{fig:hexagon-deformed}
\includegraphics[width=0.45\linewidth]{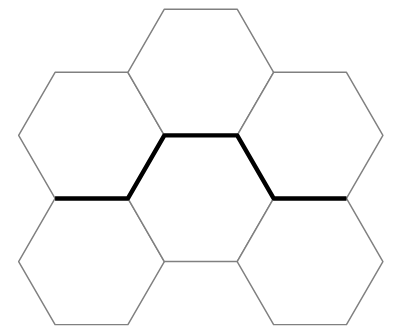}
}
    \caption{(a) The plaquette action (red) on a string excitation (black) in a hexagonal lattice contributing to the first-order corrections. The hexagon plaquette shares half of its six edges with the string. 
    (b) The resulting deformed string state. }
    \label{fig:string-hexagonal}
\end{figure}

From this dimension counting, one can see that another lattice geometry, namely a hexagonal lattice, would be even more efficient using this language. 
The string excitation on a hexagonal lattice can be described by using two letters specifying the relative directions, left ($\textsf l$) or right ($\textsf r$), i.e. it can be described as a spin-$1/2$ chain. 
The Type~C deformation relevant for the first-order correction for a hexagonal lattice switches the excitations on three consecutive edges of a hexagon plaquette to the other three (Fig.~\ref{fig:string-hexagonal}). 
This corresponds to an action on four spin sites, for example $\textsf{lrrl}$ becomes $\textsf{rllr}$ and indeed, this is the only 
action of the plaquette that results in an allowed configuration. 
Counting the number of the dimensions for the plaquette operators, it can be represented as the $2^4 = 16$-dimensional operator or 4 qubits for a hexagonal lattice while it needs a $27$ dimensional Hilbert space or 6 qubits for a square lattice. 
This observation is interesting since in the original language, the string words in a hexagonal lattice need one more auxiliary letter from the equivalent configurations in a square lattice, as the hexagonal lattice can be obtained from the square lattice by ``point-splitting" and adding an additional lattice direction between them just as the spin network setups ~\cite{Robson:1981ws}. 
The best thing one can do is projecting out the space involving the dynamics of the auxiliary letter, so the computational resource for the original language is bounded below by the minimum resource required for the square lattice. Studying the haxagonal lattice strings in  detail is beyond the scope of the present work. We are also studying those lattices in future work.

\section{Discussion and future directions\label{sec:discussion}}
In this work, we have extended the program started in~\cite{Berenstein:2023lgo}.
We showed that the first-order perturbation theory corrections to single string states at strong coupling are closed within the subsectors of fixed length, and these string states can be specified by the combination of the letters (directions in the lattice graph) that describes the path the string takes. 
The actions of the plaquette operators are merely manipulations of the letters. 

The Hamiltonian gives a nearest neighbor Hamiltonian, with additional projectors that enforce the zigzag symmetry. The goal of this paper has been to understand the features of this spin chain Hamiltonian in more detail, especially by studying subsectors with fewer letters than the generic type of path.

The two- and three-letter sectors in $2+1$-D are integrable as shown previously  in~\cite{Berenstein:2023lgo} being equivalent to free fermions. At leading order the differences in dynamics between the two-letter strings and three-letter strings in $2+1$-D are merely lattice artifacts.

We also argued here that the most general four letter sectors contain inelastic scattering events, as well as violations of the (boundary) Yang-Baxter equations,  indicating that they are likely  not-integrable. To leading order they have a rich spectrum of superselection sectors classified by a series of quantum numbers that we called knotiness and secondary knotiness. These describe the zigs and zags of the paths (the structure of the changes of
directions) and how these are intertwined.
 
It is expected that at higher orders  the dynamics will exhibit the same continuum physics (universality class) as one extrapolates from these sectors towards the continuum, and that should make the lattice system restore the rotational symmetry. 

 Higher order computations are necessary so that we can start seeing that the string motion is fully relativistic and quantizes a Nambu-Goto string worldsheet.
 
We also did not discuss the closed string states, which can also be called glueballs. These are interesting in their own right. 

We discussed the extension of the open string setup to $3+1$-D, where we saw that an integrable structure appears also in a particular three-letter subsector.  This sector does not require  projections due to the zigzag constraints, but more generic sectors are not integrable.
At this point, we can make a general statement that \textit{the introduction of the zigzag constraints violates the spin chain integrability}, except in the special case of the three-letter sectors in $2+1$-D where a basis change is available to discard such projections. 
A mathematically rigorous statement of non-integrability is still missing, which we leave for future work. 

The present work also includes a reformulation of the spin chain in another language (changes of directions instead of a fixed set of directions for links) to describe the string configurations.
It has an advantage over the more naive formulation because it already encodes the constraints, so it needs no explicit projection in the Hamiltonian to achieve the zigzag constraints. In that sense, the Hamiltonian density covers fewer sites.

This is important because if one where to try to simulate such a spin chain on a quantum computer, it requires fewer computational resources to simulate, at least to leading order. At the same time, the physical interpretation of the Hamiltonian action on the words is  less obvious and needs to be read carefully from the moves induced by the plaquettes. 
We briefly explored the possibility of using a different lattice geometry 
for $2+1$ dimensions, namely,  a hexagonal grid. The hexagonal grid in full seems to require fewer  computational quantum resources to time evolve the system in $2+1$ dimensions relative to the  square grid. 
Of course, one has to carefully evaluate which description and lattice geometry achieves in practice the minimal computational resources for a given hardware configuration, and for each Trotter step. 
We leave the development of such efficient quantum algorithms and the estimation of their computational complexity for future work.

It is really important to understand the continuum physics carefully.  This requires understanding  the approach to the roughening transition from strong coupling in detail. These computations require finding a way to systematically compute the higher-order corrections and also finding an efficient description of the glueball states. 

Since we provided a systematic translation from a string configuration to a word consisting of finite species of letters, constructing spin chains for the planar higher-order corrections should be straightforward, except for the following subtlety: given that the higher-order contributions are due to states with different nonperturbative energies, i.e. string lengths, then the perturbation takes us outside of the Hilbert space. One needs to integrate out the changes of length to go to an effective description with next to nearest neighbor corrections (plus projectors), etc. 
At least in principle, one should be able to parametrize such effects with improved lattice Hamiltonians on the spin chain (or the bulk).
The next step in the program we are outlining would be to calculate the corrected spin chain in full generality with not only {\em nearest neighbor} terms, but also next to nearest neighbor terms, even if restricted just to the  simplest 2 letter system n $2=1$ dimensions and $3+1$ dimensions, or the three letter setup, where we expect to start seeing violations of secondary knotiness induced by the perturbations in the $2+1$ problem. This three letter problem is also required to better understand the roughening transition and the restoration of rotational invariance better.

Another way to do possibly do things is to take into account variations of spin chain setups that allow clever ways to vary the length, as in ~\cite{Berenstein:2005fa}. 
One must also be  careful with the possible non-planar contributions from the lower-order corrections spoiling a naive analysis.
Such non-planar contributions can involve scattering with glueballs (closed paths). As stated before, the glueballs  are also independently interesting objects to be studied. 

Notice that the open strings we considered in the present work have fixed quark endpoints and hence the center of mass (COM) degrees of freedom have no dynamics. This is not the case for the closed string  glueball states.
Translations of glueballs are independent states, so a passage to spin chains must be more involved as it requires an extra Hilbert space to keep track of the center of mass degrees of freedom. Moreover, that Hilbert space of translates must also   interact nontrivially with the Hilbert space for the local edge configurations. This is required so that different glueball states have a different dispersion relation (mass in the sense of the center of mass motion). 

\begin{acknowledgments}
We would like to thank Richard Brower, David Gross, Marius de Leeuw and Ana Retore for many discussions. 
We would like to thank Hosho Katsura for helpful comments and generously sharing their notes with us. 
 D.B. would like to thank the Institute of Physics
at the University of Amsterdam. for their hospitality while this work was being carried out. The work of D.B. was
supported in part by the Department of Energy under grant DE-SC 0011702. 
\end{acknowledgments}

\appendix 

\section{Energy change by the plaquette action type~B\label{sec:deform-energies}}
Here, we calculate the nonperturbed energy of the state generated by a single plaquette action on a string state sharing one link as depicted in Fig.~\ref{subfig:plaq-same-shared}. 
The original string state excited along the open path $\Gamma$ with length $L$ is generated by the product of the gluon link operators acting on the vacuum state $\ket \Omega$: 
\begin{align}
    \ket \Gamma = \prod_{\ell \in \Gamma} U_\ell \ket \Omega
\end{align}
Let us call the plaquette we are acting $P$ and the shared link $m$. 
Every link operator excited in the resulting state $\ket{\Gamma'} = U_P \prod_{\ell \in \Gamma} U_\ell \ket \Omega$, except those acting on the link $m$, contributes in the same manner to the nonperturbed energy as Eq.~\eqref{eq:string-energy}, i.e. each contributes $\frac{\lambda}{2}$, and hence overall the energy of $\frac{\lambda}{2} (L + 2)$ come from these non-shared operators. 
The only nontrivial contribution comes from $U_m U_m$ in the product. 
Denoting $U_m$ as $U$ to simplify the notations and writing the fundamental indices explicitly from now on, the nonzero contributions to the nonperturbed energy are 
\begin{align}\label{eq:typeb-energy}
    &K U_{ij} U_{kl} \ket \Omega
    = \frac{\lambda}{2N} E^a E^a U_{ij} U_{kl} \ket \Omega 
    = \left(\lambda \frac{N^2 -2}{N^2}U_{ij}U_{kl} + \frac{\lambda}{2N}U_{il}U_{jk}\right)\ket \Omega 
\end{align}
Even though we have the second term which generates another string state than the state $\ket{\Gamma'}$, namely the state excited along a single string with a ``twist" at $m$, it is suppressed for large $N$. 
This means that only the first term survives in the large $N$ limit, and the generated state $\ket{\Gamma'}$ by the single plaquette action stays as an eigenstate of the nonperturbed Hamiltonian. 
The plaquette acted state lives in the energy level of four units higher 
than the original string state. 
The same mechanism occurs in the cases of the single plaquettes sharing more than one link with the original state. 
It always raises the nonperturbed energy by four units in the leading order, and hence they do not contribute to the first-order corrections. 
We need to take them into account in the second or higher-order corrections. 
The leading $\sim \mc O(1/N^0)$ order term gives the same correction as the vacuum correction, so the first nontrivial deviation comes from the subleading-order term in Eq.~\ref{eq:typeb-energy}. 

\section{Solving the two-letter Hamiltonian\label{sec:diagonal-dispersion}}
One can straightforwardly check that the XX model~\eqref{eq:H-XX} is equivalent to the free fermion theory using the Jordan-Wigner transformation as 
\begin{align}
    H_{\mrm{XX}} = -\frac{1}{\lambda} \sum_{k = -L/2+1}^{L/2} \Lambda_k \hat a_k^\dagger \hat a_k
\end{align}
$\Lambda_k$ are eigenvalues of the matrix $\frac{1}{2} (\delta_{i, j+1} + \delta_{i, j-1})$ with $i, j = 1, ..., L$. 
In the thermodynamic limit $L \rightarrow \infty$, the eigenvalues are $-\Lambda_k/\lambda = -\cos (2\pi k/L)/\lambda$, which is the energy of each of the fermion modes. 
One can regard $p = 2\pi k/L\in (-\pi, \pi]$ as the momentum of the mode in this limit. 
$\hat a_k$ and $\hat a_k^\dagger$ are the fermionic annihilation and creation operators, respectively: 
\begin{align}
    \{\hat a_k, \hat a_\ell^\dagger\} = \delta_{k \ell}, \quad 
    \{\hat a_k, \hat a_\ell\} = \{\hat a_k^\dagger, \hat a_\ell^\dagger\} = 0
\end{align}
The magnetization operator of the spin chain is translated on this fermionic basis as the fermion number operator:
\begin{align}
    \hat N = \sum_{k} \hat a_k^\dagger \hat a_k
\end{align}
The ground state of the model has exactly $L/2$ of the fermions filling all the negative fermion modes up to the fermi surface $\Lambda_{k_F} = 0$ (corresponding to $p_F = \pm \pi/2$), and all the modes above the fermi surface are unoccupied. \\

The one-particle excited state in this sector has one of the fermions ``popped" from the fermi sea to right outside of the surface.
The lowest possible excited state has a fermion right below the fermi surface popped out to right above the surface, and such excitation is gapless in the thermodynamic limit. 
Let the fermion pop from $p \in [-\pi/2, \pi/2]$, then the energy gap of the state from the ground state energy is $\cos(p)/\lambda$. 
This means these one-fermion excitations have a linear dispersion around the fermi surface, $\epsilon(p) \sim |p - p_F|/\lambda$. 
Indeed, the XX model corresponds that is the complex massless free fermion theory which is a $c = 1$ CFT.

\section{Solving the three-letter Hamiltonian\label{sec:three-letter-Bethe-ansatz}}
\subsection{Algebraic Bethe ansatz}
We saw the integrability of the three-letter Hamiltonian, so one can find its energy spectrum using the Bethe ansatz method. 
Here, we use the algebraic Bethe ansatz to compute the spectrum. 
The monodromy matrix can be written component-wise in the auxiliary space as 
\begin{align}
    T_0(\mu)
    = 
    \begin{pmatrix}
        A(\mu) & B_1(\mu) & B_2(\mu) \\
        C_1(\mu) & D_{11}(\mu) & D_{12}(\mu) \\
        C_2(\mu) & D_{21}(\mu) & D_{22}(\mu)
    \end{pmatrix}. 
\end{align}
In this expression, the transfer matrix is written as $t(\mu) = A(\mu) + \sum_a D_{aa}(\mu)$. 
The action of the $A$ and $D$ matrices on the reference state (the string state consisting only $rrrrr...$) can be found by looking at the upper-left components of the $3\times 3$ blocks of the R-matrix: 
\begin{align}
    A(\mu)\ket \Omega &= (i\cosh(\mu))^L \ket \Omega , \\ 
    D_{aa}(\mu) \ket \Omega &= \sinh^L(\mu) \ket \Omega . 
\end{align}
The algebraic Bethe ansatz constructs the series of the highest-weight states by acting $B$ matrices on the reference state $\ket \Omega$. 
Each $B$ moves the state to be in the representation with one additional box in Eq.~\eqref{eq:su2-rep}. 
This corresponds to adding either one $u$ (or $\ket +$) or $d$ (or $\ket -$) to the string. 
To calculate the spectrum of the state acted by the $B$ matrices, one needs to find the algebraic structure of the $A$, $B$, and $D$ matrices. 
The first step is to think of the Yang-Baxter equation of the R-matrices supported on one (let us label it $n$) of the physical spin sites and two additional auxiliary Hilbert spaces $0$ and $0'$: 
\begin{align}
    &R_{00'}(\lambda - \mu) R_{0n}(\lambda) R_{0'n}(\mu) 
    =  R_{0'n}(\mu)R_{0n}(\lambda)R_{00'}(\lambda - \mu) . 
\end{align}
By iteratively applying this relation to all of the physical spaces $n = 1, 2, ..., L$, then one finds the so-called TTR equation: 
\begin{align}
    &R_{00'}(\lambda - \mu) T_0(\lambda) T_{0'}(\mu)
    =  T_{0'}(\mu)T_0(\lambda)R_{00'}(\lambda - \mu) . 
\end{align}
Looking at relevant matrix components of this equation, we can read off the commutation relations of the $A$, $B$, and $D$ matrices as 
\begin{align}
    A(\lambda) B_a(\mu) 
    &= - i \coth(\lambda - \mu) B_a(\mu) A(\lambda)
    + 
 i \csch(\lambda - \mu) B_a(\lambda) A(\mu), 
    \\
    B_a(\lambda) B_b(\mu) &= B_a(\mu)B_b(\lambda), 
    \\
 D_{ab}(\lambda) B_c(\mu)
 &= i \coth(\lambda - \mu) P_{bc}^{de} B_e(\mu) D_{ad}(\lambda) 
 - i \csch(\lambda - \mu)  P_{bc}^{de} B_e(\mu) B_e(\lambda) D_{ad}(\mu) ,
\end{align}
where $P^{ab}_{cd} = \delta^a_d \delta^b_c$ is a permutation operator in $\mC^2 \otimes \mC^2$. 
$B_1$ adds one $\ket +$ and $B_2$ adds one $\ket -$, so now we can think of the newly generated states by acting the series of the $B$ matrices as
\begin{align}
    \ket{\{\mu_k\}; \Psi}
    = 
    B_{a_1}(\mu_1) \cdots B_{a_n}(\mu_n) \ket{\Omega} \Psi^{a_1 ... a_n}
\end{align}
with a nonzero number of $\ket \pm$, or the states in representations with one or more Young boxes in Eq.~\eqref{eq:su2-rep}. 
$n$ is the total number of $\ket +$ and $\ket -$, or the number of Young boxes of the representation in which the state lives. 
The actions of the on-diagonal elements $A$ and $D$ on this state are
\begin{widetext}
\begin{align}
    A(\mu) \ket{\{\mu_k\}; \Psi}
    &= \prod_{k = 1}^n (-i\coth(\mu - \mu_k)) B_{a_1}(\mu_1) \cdots B_{a_n}(\mu_n) A(\mu) \ket{\Omega} \Psi^{a_1 ... a_n} + (\text{others}) \nn
    &= (i \cosh(\mu))^L \prod_{k = 1}^n (-i\coth(\mu - \mu_k)) \ket{\{\mu_k\}; \Psi} + (\text{others})
\end{align}
and 
\begin{align}\label{eq:D-bethestate}
    &D_{bb}(\mu)\ket{\{\mu_k\}; \Psi}\nn
    &= i \coth(\mu - \mu_1) P^{c_1d_1}_{b a_1}B_{d_1}(\mu_1) D_{b c_1}(\mu) B_{a_2}(\mu_2) \cdots B_{a_n}(\mu_n) \ket{\Omega} \Psi^{a_1 ... a_n} + (\text{others}) \nn
    &= \prod_{k = 1}^n (\Tilde {T}_n)^{\{d_k\}}_{\{a_k\}}  (i \coth(\mu - \mu_k)) B_{d_1}(\mu_1) B_{d_2}(\mu_2)\cdots B_{d_n} (\mu_n) D_{b c_n}(\mu) \ket{\Omega} \Psi^{a_1 ... a_n}+ (\text{others}), 
\end{align}
\end{widetext}
where $\Tilde T_n$ is the shift operator 
\begin{align}
    (\Tilde {T}_n)^{\{d_k\}}_{\{a_k\}} = P^{c_1 d_1}_{b a_1} P^{c_2 d_2}_{c_1 a_2} \cdots P^{c_{n-1} d_{n-1}}_{c_{n-2} a_{n-1}} P^{c_n d_n}_{c_{n-1} a_n}. 
\end{align}
Note that we are interested in the action of the transfer matrix (i.e. $A + D_{11} + D_{22}$) after all, so we take the contraction over the index $b = 1,2$ for the $D$ matrix implicitly. 
The $+$(others) terms are off-diagonal parts wanted to be vanishing. 
The vanishing requirements lead to the constraints on $\mu$, the so-called Bethe ansatz equations, which we will find soon later. 
Given that you take the sum over $b$ and the only nonzero $D_{ab}$ are for $a = b$ when acted on $\ket\Omega$, we can contract $b$ and $c_n$ in $\Tilde T$. 
Hence, what we have is the product of the permutation matrices $P: \mC^2 \otimes \mC^2 \rightarrow \mC^2 \otimes \mC^2 $ with $w \otimes v \mapsto v \otimes w$\footnote{$\Tilde t$ is indeed the transfer matrix of the $SU(2)$ Heisenberg chain with zero spectral parameter. One may encounter more general cases where diagonalization of the $SU(2)$ transfer matrix with nonzero spectral parameter is required.
It is why this method is called the nested Bethe ansatz; one needs to keep applying the Bethe ansatz method for the series of models with the smaller subgroup symmetries until it reaches to $SU(2)$. }: 
\begin{align}
    \Tilde t_n = P_{12}P_{23} \cdots P_{n-1 n}. 
\end{align}
So, if we take $\Psi^{a_1...a_n}$ to be an eigenvector of $T^{\{d_k\}}_{\{a_k\}}$, $\ket{\{\mu_k\}; \Psi}$ is an eigenvector of $D_{bb}$ besides taking $\lambda$ to be the solution of the Bethe ansatz equation to make the $+$(others) terms vanish. 
Since $\Tilde t$ acts as a cyclic shift operator to the right (for example $\ket{011} \mapsto \ket{101}$ for $L = 3$), it is a root of the identity operator $\Tilde t^n = I$. 
This fact implies that the eigenvalues $\Tilde \Lambda$ of $\Tilde t$ to be the roots of unity: 
\begin{align}\label{eq:eval-shift}
    &\Tilde \Lambda_n^n = 1 
    \implies \Tilde \Lambda_n = e^{i2\pi m/n} \text{ with }m = 0, 1, ..., n-1. 
\end{align}
Plugging these eigenvalues back in Eq.~\eqref{eq:D-bethestate}, we have the action of $D$ to be 
\begin{align}
    &D_{bb}(\mu)\ket{\{\mu_k\}; \Psi}
    = 
    \prod_{k = 1}^n \Tilde \Lambda_n (i \coth(\mu - \mu_k)) (\sinh(\mu))^L \ket{\{\mu_k\}; \Psi} + (\text{others}). 
\end{align}
We have the action of the transfer matrix on this state by summing the actions of the $A$ and $D$ matrices: 
\begin{align}\label{eq:eval-transfer}
     t(\mu) \ket{\{\mu_k\}, \Psi} 
     = \Lambda(\mu) \ket{\{\mu_k\}, \Psi} 
    + (\text{others}), 
\end{align}
where 
\begin{align}
    \Lambda(\mu) &= \prod_{k = 1}^n (i\coth(\mu - \mu_k))
    \left[(-1)^n(i \cosh(\mu))^L
    + \Tilde \Lambda_n(\sinh(\mu))^L \right]
\end{align}
We need to require the $+$(others) terms to vanish so that $\Lambda(\mu)$ is the eigenvalue of the transfer matrix.
This requirement is exactly the requirement of \eqref{eq:eval-transfer} to be analytic due to the analyticity of the transfer matrix. 
As mentioned, these constraints on the values of $\{\mu_k\}$ are the Bethe ansatz equations. 
The possible singularities of $\Lambda(\mu)$ are at $\mu = \mu_j$ for one of $j = 1, 2, ..., n$. 
Hence, the Bethe ansatz equations are 
\begin{align}\label{eq:BA-eqn-nested}
    0 &= \prod_{k \neq j} (i\coth(\mu_j - \mu_k))  \left((-1)^n (i \cosh(\mu_j))^L
    + \Tilde \Lambda_n
     (\sinh(\mu_j))^L\right) . 
\end{align}
The solutions of these equations are 
\begin{align}
    \mu_j = i \arccot(e^{i\pi (2\ell_j+ n - 1 -2m/n)/L})
\end{align}
with some integer $\ell_j \in \{ 0, 1, ..., L-1\}$. 
The first $\prod_{k \neq j} (i\coth(\mu_j - \mu_k))$ factor diverges exponentially at $\mu_j = \mu_k$, so it prohibits multiple rapidities to coincide. 
The states $\ket{\{\mu_k\}, \Psi}$ become the eigenstates of the transfer matrix when $\{\mu_k\}$ are these solutions of the Bethe ansatz equations. 
In that case, the energy spectrum is 
\begin{align}\label{eq:energy-spectrum}
    E_n &= 
    i\frac{d}{d\mu} \log(\frac{\Lambda(\mu)}{(i\cosh\mu)^L})\bigg|_{\mu = 0}
    \nn 
    &= i\sum_{k = 1}^n \csch(\mu_k) \sech(\mu_k)\nn 
    &= \sum_{k = 1}^n 2 \cos(\frac{\pi(2\ell_j+ n - 1 -2m/n)}{L}). 
\end{align}
And, the spectrum of the total momentum is 
\begin{align}\label{eq:momentum-specturm}
    p_n
    &= 
    i \log(\frac{\Lambda(\mu)}{(i\cosh\mu)^L})\bigg|_{\mu = 0}
    \nn
    &= 
    i \sum_{k = 1}^n \log(i\coth(\mu_k))\nn 
    &= 
    -\sum_{k = 1}^n\frac{\pi(2\ell_j+ n - 1 -2m/n)}{L}. 
\end{align}
The physical interpretation of the spectra is simple.
The state generated by acting $n$ of the $B$ matrices contain $n$ fermionic quasi-particle excitations with momentum of $p = -\frac{\pi(2\ell_j+ n - 1 -2m/n)}{L}$ and energy of $E = 2\cos p$. 
The particles may not have degenerate values of momenta as required by the Bethe ansatz equations.

\subsection{Ground state, dispersion relations, and finite size effect}
From Eq.~\eqref{eq:energy-spectrum}, the ground state has exactly half of $L$, $n = L/2$ fermionic excitations filling up to the fermi surface with zero energy. 
Each fermion has the mode energy of $\epsilon_k = -2\sin(\frac{\pi(2\ell_k -1 + 4m/L)}{L}) $. 
With $n = L/2$, the fermions fill in the region between $\ell_F \approx 0$ and $\approx L/2$. 
To be more precise, the ground-state energy is minimized with $m =0$ and the fermions fill from $\ell_1 = 1$ to $\ell_n = L/2$. 
One can take the sum of these mode energies explicitly as 
\begin{align}\label{eq:gs-energy}
    E_0 = \sum_{k = 1}^{L/2} -2\sin(\frac{\pi(2k -1)}{L}) 
    = -2\csc(\frac{\pi}{L}).
\end{align}
The first excited state has one hole in the fermi sea right below the surface and/or one fermion right above the surface. 
This is massless in the thermodynamic limit given that one can make them arbitrarily close to the surface. 
The existence of the massless states suggests that the model has the CFT as its low-energy description. 
The central charge of the theory can be found by calculating the finite-size effect of the ground state. 
The CFT calculation expects that the ground-state energy has size dependence as 
\begin{align}
    E_0 = \epsilon_0 L - \frac{\pi v_F c}{6L} + \mc O(1/L^2).
\end{align}
Here, $v_F$ is the fermi speed of the excitations. 
The energy gap above the ground state is $2\sin(\frac{2 \pi \ell_k }{L})$, so the excitation around the fermi surface indeed has, as expected from the masslessness, the linear dispersion relation $\sim 2|p - p_F|$, with which the fermi speed is determined to be $v_F = 2$. 
The expansion of Eq.~\eqref{eq:gs-energy} about $1/L$ up to the second-leading order gives the finite size effect to be of 
\begin{align}
    E_0 
    = -\frac{2}{\pi} L -\frac{\pi}{3 L }+ \mc O\left(1/L^2\right).
\end{align}
One can read from this finite-size ground state energy that the central charge of the corresponding CFT is $c = 1$. 

Once we recover the overall coupling factor $-1/2\lambda$, the fermi speed is determined to be $v_F = 1/\lambda$. 
This matches exactly with the dispersion relation of the first-order spin chain for the two-letter string case as calculated in Sec.~\ref{sec:diagonal-dispersion}.

\bibliographystyle{unsrt}
\bibliography{ref}

@article{Polyakov:1997tj,
    author = "Polyakov, Alexander M.",
    title = "{String theory and quark confinement}",
    reportNumber = "PUPT-1743",
    doi = "10.1016/S0920-5632(98)00135-2",
    journal = "Nucl. Phys. B Proc. Suppl.",
    volume = "68",
    pages = "1--8",
    year = "1998"
}

@article{Brower:2020huh,
    author = "Brower, Richard C. and Berenstein, David and Kawai, Hiroki",
    title = "{Lattice Gauge Theory for a Quantum Computer}",
    eprint = "2002.10028",
    archivePrefix = "arXiv",
    primaryClass = "hep-lat",
    doi = "10.22323/1.363.0112",
    journal = "PoS",
    volume = "LATTICE2019",
    pages = "112",
    year = "2020"
}

@article{tHooft:1973alw,
    author = "'t Hooft, Gerard",
    editor = "Taylor, J. C.",
    title = "{A Planar Diagram Theory for Strong Interactions}",
    reportNumber = "CERN-TH-1786",
    doi = "10.1016/0550-3213(74)90154-0",
    journal = "Nucl. Phys. B",
    volume = "72",
    pages = "461",
    year = "1974"
}

@article{Berenstein:2022wlm,
    author = "Berenstein, David and Kawai, Hiroki and Brower, Richard",
    title = "{U(1) fields from qubits: An approach via D-theory algebra}",
    eprint = "2201.02412",
    archivePrefix = "arXiv",
    primaryClass = "hep-th",
    doi = "10.1103/PhysRevD.110.014506",
    journal = "Phys. Rev. D",
    volume = "110",
    number = "1",
    pages = "014506",
    year = "2024"
}

@article{Beard:1997ic,
    author = "Beard, B. B. and Brower, R. C. and Chandrasekharan, S. and Chen, D. and Tsapalis, A. and Wiese, U. -J.",
    editor = "Davies, C. T. H. and Barbour, I. M. and Bowler, K. C. and Kenway, R. D. and Pendleton, B. J. and Richards, D. G.",
    title = "{D-theory: Field theory via dimensional reduction of discrete variables}",
    eprint = "hep-lat/9709120",
    archivePrefix = "arXiv",
    reportNumber = "MIT-CTP-2675",
    doi = "10.1016/S0920-5632(97)00900-6",
    journal = "Nucl. Phys. B Proc. Suppl.",
    volume = "63",
    pages = "775--789",
    year = "1998"
}

@article{Chandrasekharan:1996ih,
    author = "Chandrasekharan, S. and Wiese, U. J.",
    title = "{Quantum link models: A Discrete approach to gauge theories}",
    eprint = "hep-lat/9609042",
    archivePrefix = "arXiv",
    reportNumber = "MIT-CTP-2573, CTP-2573",
    doi = "10.1016/S0550-3213(97)00006-0",
    journal = "Nucl. Phys. B",
    volume = "492",
    pages = "455--474",
    year = "1997"
}

@article{Brower:1997ha,
    author = "Brower, R. and Chandrasekharan, S. and Wiese, U. J.",
    title = "{QCD as a quantum link model}",
    eprint = "hep-th/9704106",
    archivePrefix = "arXiv",
    reportNumber = "MIT-CTP-2623",
    doi = "10.1103/PhysRevD.60.094502",
    journal = "Phys. Rev. D",
    volume = "60",
    pages = "094502",
    year = "1999"
}

@article{Brower:2003vy,
    author = "Brower, R. and Chandrasekharan, S. and Riederer, S. and Wiese, U. J.",
    title = "{D theory: Field quantization by dimensional reduction of discrete variables}",
    eprint = "hep-lat/0309182",
    archivePrefix = "arXiv",
    doi = "10.1016/j.nuclphysb.2004.06.007",
    journal = "Nucl. Phys. B",
    volume = "693",
    pages = "149--175",
    year = "2004"
}

@article{Gross:1973ju,
    author = "Gross, D. J. and Wilczek, Frank",
    title = "{Asymptotically Free Gauge Theories - I}",
    reportNumber = "NAL-PUB-73-49-THY, FERMILAB-PUB-73-049-T",
    doi = "10.1103/PhysRevD.8.3633",
    journal = "Phys. Rev. D",
    volume = "8",
    pages = "3633--3652",
    year = "1973"
}

@article{Dubovsky:2015zey,
    author = "Dubovsky, Sergei and Gorbenko, Victor",
    title = "{Towards a Theory of the QCD String}",
    eprint = "1511.01908",
    archivePrefix = "arXiv",
    primaryClass = "hep-th",
    doi = "10.1007/JHEP02(2016)022",
    journal = "JHEP",
    volume = "02",
    pages = "022",
    year = "2016"
}

@article{Wilson:1974sk,
    author = "Wilson, Kenneth G.",
    editor = "Taylor, J. C.",
    title = "{Confinement of Quarks}",
    reportNumber = "CLNS-262",
    doi = "10.1103/PhysRevD.10.2445",
    journal = "Phys. Rev. D",
    volume = "10",
    pages = "2445--2459",
    year = "1974"
}

@article{Berenstein:2004ys,
    author = "Berenstein, David and Cherkis, Sergey A.",
    title = "{Deformations of N=4 SYM and integrable spin chain models}",
    eprint = "hep-th/0405215",
    archivePrefix = "arXiv",
    doi = "10.1016/j.nuclphysb.2004.09.005",
    journal = "Nucl. Phys. B",
    volume = "702",
    pages = "49--85",
    year = "2004"
}

@article{DiMarcantonio:2025cmf,
    author = "Di Marcantonio, Francesco and Pradhan, Sunny and Vallecorsa, Sofia and Ba{\~n}uls, Mari Carmen and Ortega, Enrique Rico",
    title = "{Roughening and dynamics of an electric flux string in a (2+1)D lattice gauge theory}",
    eprint = "2505.23853",
    archivePrefix = "arXiv",
    primaryClass = "hep-lat",
    reportNumber = "CERN-TH-2025-105",
    month = "5",
    year = "2025"
}

@article{Ipsen:2018fmu,
    author = "Ipsen, Asger C. and Staudacher, Matthias and Zippelius, Leonard",
    title = "{The one-loop spectral problem of strongly twisted $ \mathcal{N} $ = 4 Super Yang-Mills theory}",
    eprint = "1812.08794",
    archivePrefix = "arXiv",
    primaryClass = "hep-th",
    reportNumber = "HU-Mathematik-2018-11, HU-EP-18/39",
    doi = "10.1007/JHEP04(2019)044",
    journal = "JHEP",
    volume = "04",
    pages = "044",
    year = "2019"
}

@article{Retore:2021wwh,
    author = "Retore, Ana L.",
    title = "{Introduction to classical and quantum integrability}",
    eprint = "2109.14280",
    archivePrefix = "arXiv",
    primaryClass = "hep-th",
    doi = "10.1088/1751-8121/ac5a8e",
    journal = "J. Phys. A",
    volume = "55",
    number = "17",
    pages = "173001",
    year = "2022"
}

@article{Ahn:2020zly,
    author = "Ahn, Changrim and Staudacher, Matthias",
    title = "{The Integrable (Hyper)eclectic Spin Chain}",
    eprint = "2010.14515",
    archivePrefix = "arXiv",
    primaryClass = "hep-th",
    reportNumber = "HU-Mathematik-2020-01, HU-EP-20/04",
    doi = "10.1007/JHEP02(2021)019",
    journal = "JHEP",
    volume = "02",
    pages = "019",
    year = "2021"
}

@article{Gross:1973id,
    author = "Gross, David J. and Wilczek, Frank",
    editor = "Taylor, J. C.",
    title = "{Ultraviolet Behavior of Nonabelian Gauge Theories}",
    doi = "10.1103/PhysRevLett.30.1343",
    journal = "Phys. Rev. Lett.",
    volume = "30",
    pages = "1343--1346",
    year = "1973"
}

@article{Gross:1974cs,
    author = "Gross, D. J. and Wilczek, Frank",
    title = "{ASYMPTOTICALLY FREE GAUGE THEORIES. 2.}",
    doi = "10.1103/PhysRevD.9.980",
    journal = "Phys. Rev. D",
    volume = "9",
    pages = "980--993",
    year = "1974"
}

@article{Politzer:1973fx,
    author = "Politzer, H. David",
    editor = "Taylor, J. C.",
    title = "{Reliable Perturbative Results for Strong Interactions?}",
    doi = "10.1103/PhysRevLett.30.1346",
    journal = "Phys. Rev. Lett.",
    volume = "30",
    pages = "1346--1349",
    year = "1973"
}

@article{Berenstein:2002jq,
    author = "Berenstein, David Eliecer and Maldacena, Juan Martin and Nastase, Horatiu Stefan",
    title = "{Strings in flat space and pp waves from N=4 superYang-Mills}",
    eprint = "hep-th/0202021",
    archivePrefix = "arXiv",
    doi = "10.1088/1126-6708/2002/04/013",
    journal = "JHEP",
    volume = "04",
    pages = "013",
    year = "2002"
}

@article{Kogut:1974ag,
    author = "Kogut, John B. and Susskind, Leonard",
    title = "{Hamiltonian Formulation of Wilson's Lattice Gauge Theories}",
    reportNumber = "Print-74-1186 (CORNELL)",
    doi = "10.1103/PhysRevD.11.395",
    journal = "Phys. Rev. D",
    volume = "11",
    pages = "395--408",
    year = "1975"
}

@article{Berenstein:2005fa,
    author = "Berenstein, David and Correa, Diego H. and Vazquez, Samuel E.",
    title = "{Quantizing open spin chains with variable length: An Example from giant gravitons}",
    eprint = "hep-th/0502172",
    archivePrefix = "arXiv",
    reportNumber = "NSF-KITP-05-11",
    doi = "10.1103/PhysRevLett.95.191601",
    journal = "Phys. Rev. Lett.",
    volume = "95",
    pages = "191601",
    year = "2005"
}

@article{Polchinski:1991ax,
    author = "Polchinski, Joseph and Strominger, Andrew",
    title = "{Effective string theory}",
    reportNumber = "UTTG-17-91",
    doi = "10.1103/PhysRevLett.67.1681",
    journal = "Phys. Rev. Lett.",
    volume = "67",
    pages = "1681--1684",
    year = "1991"
}

@article{Kogut1981,
  title = {Fluctuating string of lattice gauge theory: The heavy-quark potential, the restoration of rotational symmetry, and roughening},
  author = {Kogut, J. B. and Sinclair, D. K. and Pearson, R. B. and Richardson, J. L. and Shigemitsu, J.},
  journal = {Phys. Rev. D},
  volume = {23},
  issue = {12},
  pages = {2945--2961},
  numpages = {0},
  year = {1981},
  month = {Jun},
  publisher = {American Physical Society},
  doi = {10.1103/PhysRevD.23.2945},
  url = {https://link.aps.org/doi/10.1103/PhysRevD.23.2945}
}

@article{Luscher:1980ac,
    author = "Luscher, M.",
    title = "{Symmetry Breaking Aspects of the Roughening Transition in Gauge Theories}",
    reportNumber = "DESY-80-87",
    doi = "10.1016/0550-3213(81)90423-5",
    journal = "Nucl. Phys. B",
    volume = "180",
    pages = "317--329",
    year = "1981"
}

@article{Juge:2001mj,
    author = "Juge, K. Jimmy and Kuti, Julius and Morningstar, Colin J.",
    editor = "Muller-Preussker, M. and Bietenholz, Wolfgang and Jansen, K. and Jegerlehner, F. and Montvay, I. and Schierholz, G. and Sommer, R. and Wolff, U.",
    title = "{The QCD String Spectrum and Conformal Field Theory}",
    eprint = "hep-lat/0110157",
    archivePrefix = "arXiv",
    reportNumber = "FERMILAB-CONF-01-316-T",
    doi = "10.1016/S0920-5632(01)01818-7",
    journal = "Nucl. Phys. B Proc. Suppl.",
    volume = "106",
    pages = "691--693",
    year = "2002"
}

@inproceedings{Juge:2001rb,
    author = "Juge, K. Jimmy and Kuti, Julius and Morningstar, Colin",
    title = "{From Surface Roughening to QCD String Theory}",
    booktitle = "{24th Johns Hopkins Workshop on Nonperturbative QFT Methods and Their Applications}",
    eprint = "hep-lat/0103008",
    archivePrefix = "arXiv",
    reportNumber = "FERMILAB-CONF-01-069-T",
    doi = "10.1142/9789812799968_0006",
    pages = "143--165",
    month = "3",
    year = "2001"
}

@article{Alcaraz:1991ps,
    author = "Alcaraz, Francisco C. and Koberle, Roland and Lima-Santos, A.",
    title = "{All exactly solvable quantum spin 1 chains from Hecke algebra}",
    reportNumber = "NSF-ITP-92-02",
    doi = "10.1142/S0217751X92003458",
    journal = "Int. J. Mod. Phys. A",
    volume = "7",
    pages = "7615--7628",
    year = "1992"
}

@article{Alcaraz:1992zc,
    author = "Alcaraz, Francisco C. and Droz, Michel and Henkel, Malte and Rittenberg, Vladimir",
    title = "{Reaction - diffusion processes, critical dynamics and quantum chains}",
    eprint = "hep-th/9302112",
    archivePrefix = "arXiv",
    reportNumber = "UGVA-DPT-1992-12-799",
    doi = "10.1006/aphy.1994.1026",
    journal = "Annals Phys.",
    volume = "230",
    pages = "250--302",
    year = "1994"
}

@article{Gomez:1992nd,
    author = "Gomez, Cesar and Sierra, German",
    title = "{New integrable deformations of higher spin Heisenberg-Ising chains}",
    eprint = "hep-th/9203035",
    archivePrefix = "arXiv",
    reportNumber = "UGVA-DPT-1992-03-760",
    doi = "10.1016/0370-2693(92)91310-6",
    journal = "Phys. Lett. B",
    volume = "285",
    pages = "126--132",
    year = "1992"
}

@article{Maassarani:1997kon,
    author = "Maassarani, Z. and Mathieu, P.",
    title = "{The su(N) XX model}",
    eprint = "cond-mat/9709163",
    archivePrefix = "arXiv",
    doi = "10.1016/S0550-3213(98)80004-7",
    journal = "Nucl. Phys. B",
    volume = "517",
    pages = "395--408",
    year = "1998"
}

@misc{Katsura2023,
    author = "Katsura, Hosho",
    title = "Private communication",
    year = "2023"
}

@article{Itoi1997,
  title = {Extended massless phase and the Haldane phase in a spin-1 isotropic antiferromagnetic chain},
  author = {Itoi, Chigak and Kato, Masa-Hide},
  journal = {Phys. Rev. B},
  volume = {55},
  issue = {13},
  pages = {8295--8303},
  numpages = {0},
  year = {1997},
  month = {Apr},
  publisher = {American Physical Society},
  doi = {10.1103/PhysRevB.55.8295},
  url = {https://link.aps.org/doi/10.1103/PhysRevB.55.8295}
}

@article{Ambjorn:1999ei,
    author = "Ambjorn, Jan and Karakhanian, D. and Mirumian, M. and Sedrakian, A.",
    title = "{Fermionization of the spin S Uimin-Lai-Sutherland model: Generalization of supersymmetric t - J model to spin S}",
    eprint = "cond-mat/9909432",
    archivePrefix = "arXiv",
    reportNumber = "NBI-HE-99-38",
    doi = "10.1016/S0550-3213(00)00757-4",
    journal = "Nucl. Phys. B",
    volume = "599",
    pages = "547--560",
    year = "2001"
}

@article{Bauer:2022hpo,
    author = "Bauer, Christian W. and others",
    title = "{Quantum Simulation for High-Energy Physics}",
    eprint = "2204.03381",
    archivePrefix = "arXiv",
    primaryClass = "quant-ph",
    reportNumber = "UMD-PP-022-04, LA-UR-22-22100, RIKEN-iTHEMS-Report-22, RIKEN-iTHEMS-Report-22,
  FERMILAB-PUB-22-249-SQMS-T, FERMILAB-PUB-22-249-SQMS-T, IQuS@UW-21-027, MITRE-21-03848-2, FERMILAB-PUB-22-249-SQMS-T",
    doi = "10.1103/PRXQuantum.4.027001",
    journal = "PRX Quantum",
    volume = "4",
    number = "2",
    pages = "027001",
    year = "2023"
}

@article{Berenstein:2023lgo,
    author = "Berenstein, David and Kawai, Hiroki",
    title = "{Integrable Spin Chains from large-$N$ QCD at strong coupling}",
    eprint = "2308.11716",
    archivePrefix = "arXiv",
    primaryClass = "hep-th",
    month = "8",
    year = "2023"
}

@article{Rinaldi:2021jbg,
    author = "Rinaldi, Enrico and Han, Xizhi and Hassan, Mohammad and Feng, Yuan and Nori, Franco and McGuigan, Michael and Hanada, Masanori",
    title = "{Matrix-Model Simulations Using Quantum Computing, Deep Learning, and Lattice Monte Carlo}",
    eprint = "2108.02942",
    archivePrefix = "arXiv",
    primaryClass = "quant-ph",
    reportNumber = "RIKEN-iTHEMS-Report-21, DMUS-MP-21/10",
    doi = "10.1103/PRXQuantum.3.010324",
    journal = "PRX Quantum",
    volume = "3",
    number = "1",
    pages = "010324",
    year = "2022"
}

@article{Banks:1996vh,
    author = "Banks, Tom and Fischler, W. and Shenker, S. H. and Susskind, Leonard",
    title = "{M theory as a matrix model: A conjecture}",
    eprint = "hep-th/9610043",
    archivePrefix = "arXiv",
    reportNumber = "RU-96-95, SU-ITP-96-12, UTTG-13-96",
    doi = "10.1201/9781482268737-37",
    journal = "Phys. Rev. D",
    volume = "55",
    pages = "5112--5128",
    year = "1997"
}

@article{Ciavarella:2024fzw,
    author = "Ciavarella, Anthony N. and Bauer, Christian W.",
    title = "{Quantum Simulation of SU(3) Lattice Yang-Mills Theory at Leading Order in Large-Nc Expansion}",
    eprint = "2402.10265",
    archivePrefix = "arXiv",
    primaryClass = "hep-ph",
    doi = "10.1103/PhysRevLett.133.111901",
    journal = "Phys. Rev. Lett.",
    volume = "133",
    number = "11",
    pages = "111901",
    year = "2024"
}

@article{Byrnes:2005qx,
    author = "Byrnes, Tim and Yamamoto, Yoshihisa",
    title = "{Simulating lattice gauge theories on a quantum computer}",
    eprint = "quant-ph/0510027",
    archivePrefix = "arXiv",
    doi = "10.1103/PhysRevA.73.022328",
    journal = "Phys. Rev. A",
    volume = "73",
    pages = "022328",
    year = "2006"
}

@article{Ciavarella:2025bsg,
    author = "Ciavarella, Anthony N. and Burbano, Ivan M. and Bauer, Christian W.",
    title = "{Efficient truncations of SU(Nc) lattice gauge theory for quantum simulation}",
    eprint = "2503.11888",
    archivePrefix = "arXiv",
    primaryClass = "hep-lat",
    doi = "10.1103/ylqb-phv5",
    journal = "Phys. Rev. D",
    volume = "112",
    number = "5",
    pages = "054514",
    year = "2025"
}

@article{Kan:2021xfc,
    author = "Kan, Angus and Nam, Yunseong",
    title = "{Lattice Quantum Chromodynamics and Electrodynamics on a Universal Quantum Computer}",
    eprint = "2107.12769",
    archivePrefix = "arXiv",
    primaryClass = "quant-ph",
    month = "7",
    year = "2021"
}

@article{Santra:2025dsm,
    author = "Santra, Gopal Chandra and Mildenberger, Julius and Ballini, Edoardo and Bottarelli, Alberto and Wauters, Matteo M. and Hauke, Philipp",
    title = "{Quantum Resources in Non-Abelian Lattice Gauge Theories: Nonstabilizerness, Multipartite Entanglement, and Fermionic Non-Gaussianity}",
    eprint = "2510.07385",
    archivePrefix = "arXiv",
    primaryClass = "quant-ph",
    month = "10",
    year = "2025"
}

@article{Raychowdhury:2019iki,
    author = "Raychowdhury, Indrakshi and Stryker, Jesse R.",
    title = "{Loop, string, and hadron dynamics in SU(2) Hamiltonian lattice gauge theories}",
    eprint = "1912.06133",
    archivePrefix = "arXiv",
    primaryClass = "hep-lat",
    reportNumber = "INT-PUB-19-059, UMD-PP-019-08",
    doi = "10.1103/PhysRevD.101.114502",
    journal = "Phys. Rev. D",
    volume = "101",
    number = "11",
    pages = "114502",
    year = "2020"
}

@article{Kadam:2022ipf,
    author = "Kadam, Saurabh V. and Raychowdhury, Indrakshi and Stryker, Jesse R.",
    title = "{Loop-string-hadron formulation of an SU(3) gauge theory with dynamical quarks}",
    eprint = "2212.04490",
    archivePrefix = "arXiv",
    primaryClass = "hep-lat",
    reportNumber = "UMD-PP-022-11",
    doi = "10.1103/PhysRevD.107.094513",
    journal = "Phys. Rev. D",
    volume = "107",
    number = "9",
    pages = "094513",
    year = "2023"
}

@article{Zache:2023dko,
    author = "Zache, Torsten V. and Gonz{\'a}lez-Cuadra, Daniel and Zoller, Peter",
    title = "{Quantum and Classical Spin-Network Algorithms for q-Deformed Kogut-Susskind Gauge Theories}",
    eprint = "2304.02527",
    archivePrefix = "arXiv",
    primaryClass = "quant-ph",
    doi = "10.1103/PhysRevLett.131.171902",
    journal = "Phys. Rev. Lett.",
    volume = "131",
    number = "17",
    pages = "171902",
    year = "2023"
}

@article{Hayata:2023bgh,
    author = "Hayata, Tomoya and Hidaka, Yoshimasa",
    title = "{q deformed formulation of Hamiltonian SU(3) Yang-Mills theory}",
    eprint = "2306.12324",
    archivePrefix = "arXiv",
    primaryClass = "hep-lat",
    reportNumber = "KEK-TH-2533, J-PARC-TH-0291, RIKEN-iTHEMS-Report-23",
    doi = "10.1007/JHEP09(2023)123",
    journal = "JHEP",
    volume = "09",
    pages = "123",
    year = "2023"
}

@article{Zohar:2016iic,
    author = "Zohar, Erez and Farace, Alessandro and Reznik, Benni and Cirac, J. Ignacio",
    title = "{Digital lattice gauge theories}",
    eprint = "1607.08121",
    archivePrefix = "arXiv",
    primaryClass = "quant-ph",
    doi = "10.1103/PhysRevA.95.023604",
    journal = "Phys. Rev. A",
    volume = "95",
    number = "2",
    pages = "023604",
    year = "2017"
}

@article{Lamm:2019bik,
    author = "Lamm, Henry and Lawrence, Scott and Yamauchi, Yukari",
    title = "{General Methods for Digital Quantum Simulation of Gauge Theories}",
    eprint = "1903.08807",
    archivePrefix = "arXiv",
    primaryClass = "hep-lat",
    doi = "10.1103/PhysRevD.100.034518",
    journal = "Phys. Rev. D",
    volume = "100",
    number = "3",
    pages = "034518",
    year = "2019"
}

@article{Gross:1993hu,
    author = "Gross, David J. and Taylor, Washington",
    title = "{Two-dimensional QCD is a string theory}",
    eprint = "hep-th/9301068",
    archivePrefix = "arXiv",
    reportNumber = "LBL-33458, PUPT-1376, UCB-PTH-93-02",
    doi = "10.1016/0550-3213(93)90403-C",
    journal = "Nucl. Phys. B",
    volume = "400",
    pages = "181--208",
    year = "1993"
}

@article{Berenstein:2006qk,
    author = "Berenstein, David and Correa, Diego H. and Vazquez, Samuel E.",
    title = "{A Study of open strings ending on giant gravitons, spin chains and integrability}",
    eprint = "hep-th/0604123",
    archivePrefix = "arXiv",
    reportNumber = "CECS-PHY-06-07",
    doi = "10.1088/1126-6708/2006/09/065",
    journal = "JHEP",
    volume = "09",
    pages = "065",
    year = "2006"
}

@article{Modi:2026syn,
    author = "Modi, Neel S. and Ciavarella, Anthony N. and Halimeh, Jad C. and Bauer, Christian W.",
    title = "{Large Nc Truncations for SU(Nc) Lattice Yang-Mills Theory with Fermions}",
    eprint = "2602.02344",
    archivePrefix = "arXiv",
    primaryClass = "hep-lat",
    month = "2",
    year = "2026"
}

@article{Robson:1981ws,
    author = "Robson, D. and Webber, D. M.",
    title = "{Gauge Covariance in Lattice Field Theories}",
    reportNumber = "LTH 75",
    doi = "10.1007/BF01475006",
    journal = "Z. Phys. C",
    volume = "15",
    pages = "199",
    year = "1982"
}

@article{Wong:1994np,
    author = "Wong, E. and Affleck, I.",
    title = "{Tunneling in quantum wires: A Boundary conformal field theory approach}",
    eprint = "cond-mat/9311040",
    archivePrefix = "arXiv",
    doi = "10.1016/0550-3213(94)90479-0",
    journal = "Nucl. Phys. B",
    volume = "417",
    pages = "403--438",
    year = "1994"
}

\end{document}